\documentclass[aps,prd,reprint,preprintnumbers,superscriptaddress,showpacs,twocolumn]{revtex4-1}
\usepackage{latexsym,graphicx,amssymb,amsmath,mathrsfs}
\usepackage{setspace,bm}
\usepackage[breaklinks, colorlinks=true, pdfstartview=FitV, linkcolor=red, citecolor=blue, urlcolor=blue]{hyperref}
\usepackage[usenames]{color}
\usepackage{latexsym}
\usepackage{epstopdf}
\usepackage{mathtools}
\usepackage{url}
\usepackage{comment}
\usepackage{braket}
\usepackage{CJKutf8}

\newcommand{\bequ}{\begin{equation}}
\newcommand{\eequ}{\end{equation}}
\newcommand{\bea}{\begin{eqnarray}}
\newcommand{\eea}{\end{eqnarray}}

\newcommand{\m}{\mu}

\DeclareSymbolFont{boldletters}{OML}{cmm} {b}{it}
\DeclareSymbolFontAlphabet{\mathbit}{boldletters}
\DeclareMathSymbol{\alpha}{\mathalpha}{letters}{"0B}
\DeclareMathSymbol{\beta}{\mathalpha}{letters}{"0C}
\DeclareMathSymbol{\gamma}{\mathalpha}{letters}{"0D}
\DeclareMathSymbol{\delta}{\mathalpha}{letters}{"0E}
\DeclareMathSymbol{\epsilon}{\mathalpha}{letters}{"0F}
\DeclareMathSymbol{\zeta}{\mathalpha}{letters}{"10}
\DeclareMathSymbol{\eta}{\mathalpha}{letters}{"11}
\DeclareMathSymbol{\theta}{\mathalpha}{letters}{"12}
\DeclareMathSymbol{\iota}{\mathalpha}{letters}{"13}
\DeclareMathSymbol{\kappa}{\mathalpha}{letters}{"14}
\DeclareMathSymbol{\lambda}{\mathalpha}{letters}{"15}
\DeclareMathSymbol{\mu}{\mathalpha}{letters}{"16}
\DeclareMathSymbol{\nu}{\mathalpha}{letters}{"17}
\DeclareMathSymbol{\xi}{\mathalpha}{letters}{"18}
\DeclareMathSymbol{\pi}{\mathalpha}{letters}{"19}
\DeclareMathSymbol{\rho}{\mathalpha}{letters}{"1A}
\DeclareMathSymbol{\sigma}{\mathalpha}{letters}{"1B}
\DeclareMathSymbol{\tau}{\mathalpha}{letters}{"1C}
\DeclareMathSymbol{\upsilon}{\mathalpha}{letters}{"1D}
\DeclareMathSymbol{\phi}{\mathalpha}{letters}{"1E}
\DeclareMathSymbol{\chi}{\mathalpha}{letters}{"1F}
\DeclareMathSymbol{\psi}{\mathalpha}{letters}{"20}
\DeclareMathSymbol{\omega}{\mathalpha}{letters}{"21}
\DeclareMathSymbol{\varepsilon}{\mathalpha}{letters}{"22}
\DeclareMathSymbol{\vartheta}{\mathalpha}{letters}{"23}
\DeclareMathSymbol{\varpi}{\mathalpha}{letters}{"24}
\DeclareMathSymbol{\varrho}{\mathalpha}{letters}{"25}
\DeclareMathSymbol{\varsigma}{\mathalpha}{letters}{"26}
\DeclareMathSymbol{\varphi}{\mathalpha}{letters}{"27}
\DeclareMathSymbol{\Gamma}{\mathalpha}{letters}{"00}
\DeclareMathSymbol{\Delta}{\mathalpha}{letters}{"01}
\DeclareMathSymbol{\Theta}{\mathalpha}{letters}{"02}
\DeclareMathSymbol{\Lambda}{\mathalpha}{letters}{"03}
\DeclareMathSymbol{\Xi}{\mathalpha}{letters}{"04}
\DeclareMathSymbol{\Pi}{\mathalpha}{letters}{"05}
\DeclareMathSymbol{\Sigma}{\mathalpha}{letters}{"06}
\DeclareMathSymbol{\Upsilon}{\mathalpha}{letters}{"07}
\DeclareMathSymbol{\Phi}{\mathalpha}{letters}{"08}
\DeclareMathSymbol{\Psi}{\mathalpha}{letters}{"09}
\DeclareMathSymbol{\Omega}{\mathalpha}{letters}{"0A}

\begin{document}
\title{Thermodynamic geometry in hadron resonance gas model at real and imaginary baryon chemical potential and a simple sufficient condition for quark deconfinement}

\author{Riki Oshima}
\email[]{24804001@edu.cc.saga-u.ac.jp}
\affiliation{Department of Physics, Saga University,
             Saga 840-8502, Japan}

\author{Hiroaki Kouno}
\email[]{kounoh@cc.saga-u.ac.jp}
\affiliation{Department of Physics, Saga University,
             Saga 840-8502, Japan}

\author{Motoi Tachibana}
\email[]{motoi@cc.saga-u.ac.jp}
\affiliation{Department of Physics, Saga University,
             Saga 840-8502, Japan}
\affiliation{Center for Theoretical Physics, Khazar University, 41 Mehseti Street, Baku, AZ1096, Azerbaijan}

\author{Kouji Kashiwa}
\email[]{kashiwa@fit.ac.jp}
\affiliation{Fukuoka Institute of Technology, Wajiro, Fukuoka 811-0295, Japan}


\begin{abstract}
The thermodynamic geometry of the hadron resonance gas model with (without) excluded volume effects (EVE) of baryons is investigated.  
The case with imaginary $\mu$, where $\mu$ is the baryon chemical potential, is investigated as well as the one with real $\mu$. 
We calculate the scalar curvature $R$ and use the $R=0$ criterion to investigate the phase structure in the $\mu^2$-$T$ plane where $T$ is the temperature.  
The curve on which $R=0$ continues analytically from the imaginary $\mu$ region, where the lattice QCD is feasible,  to the real $\mu$ one. 
In the presence of EVE, there are rich phase structures in the large real $\mu$ region as well as the Roberge-Weiss like region where $\mu$ is imaginary and a singularity appears, while there is no phase structure in the large real $\mu$ region in the absence of EVE. 
The limiting temperature of the baryon gas is also obtained by using the baryon number fluctuation.  
The LQCD predicted critical point locates almost on the curve of the limiting temperature we determined. 
A simple empiric sufficient condition, $n_{\rm B}>1/(2v_{\rm B})$, is obtained for the quark deconfinement in the large real $\mu$ region, where $n_{\rm B}$ and $v_{\rm B}$ are the net baryon number density and the volume of a baryon, respectively.  
\end{abstract}

\maketitle


\section{Introduction}

The determination of the phase diagram of quantum chromodynamics (QCD) is an important subject not only in nuclear and particle physics but also in cosmology and astrophysics; see, e.g., Ref.~\cite{Fukushima:2010bq} and references therein.  
However, when the baryon number chemical potential is finite and real, the first-principle calculation, namely, the lattice QCD (LQCD) simulation, is not feasible due to the infamous sign problem; see Refs.\,\cite{deForcrand:2010ys,Nagata:2021bru,*Nagata:2021ugx} as an example. 
To circumvent the sign problem, several methods have been proposed and investigated, although, at present, these methods are not complete and we do not have adequate information on the equation of state (EOS) at finite baryon density. 
 
So far, the properties of the nuclear/quark matter at high baryon density are mainly investigated by using effective models of QCD. 
It is known that the LQCD results at $\mu = 0$ are in good agreement with those obtained by the hadron resonance gas (HRG) model when temperature $T$ is not so large.  
Usually, the ideal gas approximation is used for the calculations in the HRG model.  
However, it is expected that repulsive effects among baryons are important at high density.  
If the repulsion is absent, baryon matter is realized at a sufficiently large baryon density~\cite{Cleymans:1985wb}. 
One of the traditional treatments for such repulsion is to consider excluded volume effects (EVE) among baryons~\cite{Cleymans:1986cq, Kouno:1988bi, Rischke:1991ke}.   
EVE successfully prevents the realization of baryon matter at large baryon chemical potential~\cite{Cleymans:1986cq}; for the recent review, see, e.g., Ref.~\cite{Fujimoto:2021dvn} and references therein.
The availability of the HRG model with EVE may be checked by using the LQCD results at finite imaginary $\mu$.  

When the baryon (quark) chemical potential $\mu$ ($\mu_{\rm q}$) is pure imaginary, no sign problem occurs and one can perform LQCD simulations; see Refs.\,\cite{deForcrand:2002ci,D'Elia:2002gd,D'Elia:2004at,Chen:2004tb,D'Elia:2009qz} as an example. 
The grand canonical QCD partition function $Z(\theta)$ with pure imaginary quark chemical potential ($\mu_{\rm q}={\mu / 3}=i\theta_{\rm q} T$) has the Roberge-Weiss (RW) periodicity~\cite{Roberge:1986mm} as
\begin{eqnarray}
Z \Bigl(\theta_{\rm q} + 
       {2\pi\over{3}} \Bigr) 
= Z(\theta_{\rm q}), 
\label{RWP1}
\end{eqnarray}
where $T$ is the temperature and $\theta_\mathrm{q} \in \mathbb{R}$. 
This periodicity is the remnant of the $\mathbb{Z}_3$-symmetry of pure gluon theory. 
At low temperature, $Z(\theta_{\rm q} )$ is expected to be a smooth function of $\theta_{\rm q}$. 
However, at high temperature above the RW temperature $T_{\rm RW}$, it has a singularity at $\theta_{\rm q}=(2k+1) \pi / 3$ where $k \in \mathbb{Z}$.   
This singularity is called the RW transition. 
$T_{\rm RW}$ for 2+1 flavor QCD is estimated as about 200~MeV by LQCD simulations~\cite{Bonati:2016pwz,Cuteri:2022vwk,Bonati:2018fvg}. 

In the HRG model with pure imaginary baryon number chemical potential $\mu =i\theta T~(=i3\theta_{\rm q}T)$, the RW periodicity is trivial since the model has a trivial periodicity
\begin{eqnarray}
Z_{\rm HRG}(\theta  +2\pi )=Z_{\rm HRG}(\theta). 
\label{RWP_HRG}
\end{eqnarray}
In the case of the free hadron resonance gas model, $Z_{\rm HRG}(\theta )$ is a smooth function of $\theta$ at any temperature. 
However, it has a singularity when interaction effects such as EVE are taken into account~\cite{Taradiy:2019taz,Savchuk:2019yxl,Vovchenko:2017xad}.
In Ref.~\cite{Oshima:2023bip}, the $\theta$-dependence of the baryon number density and the pressure was studied in detail. 
It was found that these quantities have singularities at $\theta =\pi$ when $T\sim 210$~MeV. 
We call this temperature the Roberge-Weiss-like (RWL) temperature $T_{\rm RWL}$ in this paper.  

\begin{figure}[t]
\centering
\centerline{\includegraphics[width=0.40\textwidth]{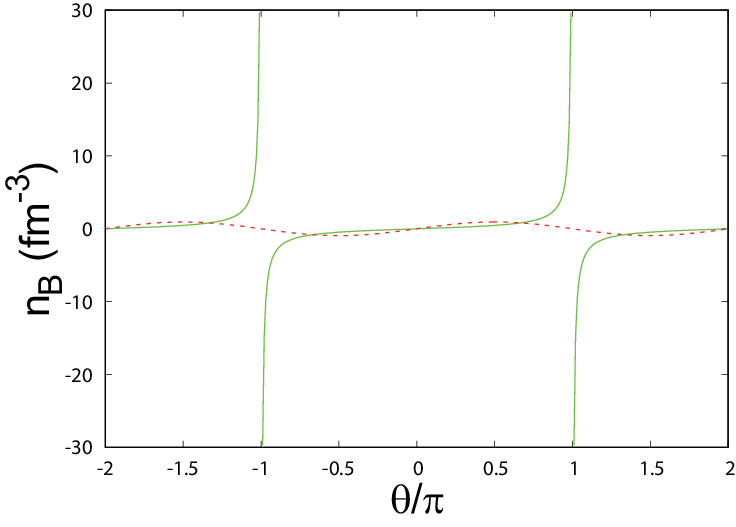}}
\caption{The $\theta$-dependence of the net baryon number density $n_{\rm B}$ in the HRG model, when $T=T_{\rm RWL}=0.2103$~GeV.  
The solid and dashed lines show the results with EVE and without EVE, respectively. 
For the details of the calculations, see Sec.~\ref{HRGEVE}. 
}
 \label{Fig_n_TRW_theta}
\end{figure}

\begin{figure}[h]
\centering
\centerline{\includegraphics[width=0.40\textwidth]{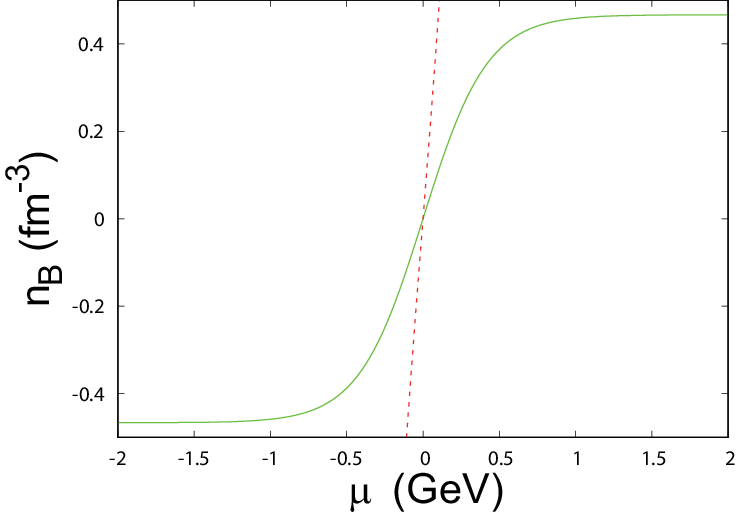}}
\caption{The $\mu$-dependence of the net baryon number density $n_{\rm B}$ in the HRG model, when $T=T_{\rm RWL}=0.2103$~GeV. 
The solid and dashed lines show the results with EVE and without EVE, respectively. 
For the details of the calculations, see Sec.~\ref{HRGEVE}. 
}
 \label{Fig_n_TRW_mu}
\end{figure}

Figure~\ref{Fig_n_TRW_theta} shows the $\theta$-dependence of the net baryon density $n_{\rm B}$ at $T=T_{\rm RWL}$. 
In the case of the HRG model with EVE, $n_{\rm B}$ presents a discontinuous singularity at $\theta =\pm\pi$, while it is smooth function of $\theta$ at those points in the absence of the EVE. 
Figure~\ref{Fig_n_TRW_mu} shows the \real-$\mu$ dependence of the net baryon density $n_{\rm B}$ at $T=T_{\rm RWL}$. 
In the case of the HRG model with EVE, $n_{\rm B}$ saturates to the constant value $\pm 1/v_{\rm B}$ in the limit $\mu\to \pm\infty$ where $v_{\rm B}$ is the volume of a baryon, while the absolute value of it increases rapidly in the absence of the EVE as $|\mu|$ increases. 
It is clear that the saturation in the real $\mu$ region is strongly correlated with the singularity in the imaginary $\mu$ region. 
The singular behavior in the vicinity of the RWL point contains significant information about the nuclear equation of state at high baryon density. 
However, the nature of the interactions at the RWL point is not clear yet.


One of the powerful tools to investigate the features of the interaction in the thermodynamic system is the thermodynamic geometry; for a review, see, e.g, Ref.~\cite{Ruppeiner:1995zz} and references therein. 
In the thermodynamic geometry, the scalar curvature $R$ is an important quantity. 
In particle and nuclear physics, thermodynamic geometry was used to investigate the QCD phase diagram in $\mu$-$T$ plane~\cite{Castorina:2018ayy}. 
It was found that the phase diagram obtained by the thermodynamic geometry in the HRG model with EVE is consist with the pseudo-critical temperature obtained in LQCD~\cite{Castorina:2018gsx}. 
The excluded volume effects are important for reproducing the pseudo-critical temperature. 

In this paper, to investigate the nature of the EVE in detail, we study the HRG model with/without EVE at imaginary $\mu$ as well as real $\mu$ using the thermodynamic geometry. 
This paper is organized as follows. 
In Sec.~\ref{RWP}, the RW periodicity and transition are briefly reviewed.  
In Sec.~\ref{HRGEVE}, we show our formulation of the HRG model with EVE. 
In Sec.~\ref{TG}, we briefly review the thermodynamic geometry. 
In Sec. \ref{Scurve}, the numerical results of the scalar curvature and the phase structures are shown.  
In Sec. \ref{LTbaryon}, we investigate the limiting temperature of our HRG model with EVE. 
Using the limiting temperature, an approximate location of the critical point and a simple empiric sufficient condition for quark deconfinement are obtained. 
The baryon radius dependence of the results is discussed in Sec.~\ref{Baryonradius}. 
Section \ref{summary} is devoted to the summary and discussions.

\section{Roberge-Weiss periodicity and transition}
\label{RWP}

The grand canonical partition function of QCD with imaginary quark chemical potential $\mu_{\rm q} =i\theta_{\rm q}T$ is given by
\begin{align}
Z(\theta_{\rm q})
&= \int {\cal D} \psi {\cal D} \bar{\psi} {\cal D} A_\mu e^{-S(\theta_{\rm q} )}, 
\label{Z-QCD}
\end{align}
where the action is defined as
\begin{align}
S(\theta_{\rm q})
&=\int_0^\beta d\tau\int_{-\infty}^\infty d^3x \, {\cal L}(\theta_{\rm q}),
\label{S-QCD}
\end{align}
with the Lagrangian density 
\begin{align}
{\cal L} (\theta_{\rm q})
&= \bar{\psi}(\gamma_\mu D_\mu -m_0)\psi 
 -{1\over{4}}F_{\mu\nu}^2-i{\theta_{\rm q}\over{\beta}}\bar{\psi}\gamma_4\psi. 
\label{L-QCD}
\end{align}
Here $\psi$, $A_\mu$, $F_{\mu\nu}$, $D_\mu$,  and $m_0$ are the quark field, the gluon field, the gluon field strength, the covariant derivative and the current quark mass matrix, respectively, and $\beta = 1/T$. 
We use the Euclidean notation in Eqs. (\ref{Z-QCD}) $\sim$ (\ref{L-QCD}).  

To eliminate the $\theta_{\rm q}$-dependent term from the action, we perform the following transformation of quark field, 
\begin{eqnarray}
\psi \mapsto \exp{\left(i{\tau \theta_{\rm q}\over{\beta}}\right)}\psi. 
\label{transform}
\end{eqnarray}
As a result, the anti-periodic temporal boundary condition of the quark field $\psi$ is changed into 
\begin{eqnarray}
\psi ({\bf x},\beta )=-\exp{(i\theta_{\rm q})}\psi ({\bf x},0). 
\label{quark_boundary_1}
\end{eqnarray}
Hence, $\theta_{\rm q}$ can be considered as the phase of the temporal boundary condition of the quark field.  

Next, we perform another transformation of the quark and gluon fields as follows:    
\begin{eqnarray}
A_\mu&\mapsto& U({\bf x},\tau )A_\mu U^{-1}({\bf x},\tau )-{i\over{g}}(\partial_\mu U({\bf x},\tau ))U^{-1}({\bf x},\tau ), 
\nonumber\\
\psi &\mapsto & U({\bf x},\tau )\psi, 
\label{Z3trans}
\end{eqnarray}
where $g$ is a coupling constant. 
Here $U({\bf x},\tau)$ is the $SU(3)$ element which satisfies the temporal boundary condition $U({\bf x},\beta)=z_{3,k}U({\bf x},0)$ where $z_{3,k}$ is the $\mathbb{Z}_3$ element $z_{3,k}=\exp(i2\pi k/3)$ with any integer $k$.    
Although the action $S(\theta_{\rm q})$ is invariant under this ${\mathbb Z}_3$  transformation,  the boundary condition (\ref{quark_boundary_1}) of the quark field is changed into  
\begin{eqnarray}
\psi ({\bf x},\beta )=-\exp{\Big[i\Big( \theta_{\rm q} +{2\pi k\over{3}}\Big)\Big]}\psi ({\bf x},0). 
\label{quark_boundary}
\end{eqnarray}
Under this $\mathbb{Z}_3$ transformation, $Z(\theta_{\rm q})$ is changed into $Z\left(\theta_{\rm q} +{2\pi k\over{3}}\right)$.
Hence, we obtain the RW periodicity ~\cite{Roberge:1986mm}; 
\begin{eqnarray}
Z\left(\theta_{\rm q} +{2\pi k\over{3}}\right)=Z(\theta_{\rm q}). 
\label{period_RW}
\end{eqnarray}
The pure gluon theory has the $\mathbb{Z}_3$-symmetry. 
The introduction of the dynamical quarks into the theory breaks the symmetry but the RW periodicity appears as a remnant of the $\mathbb{Z}_3$-symmetry.

When $T<T_{\rm RW}$ where $T_{\rm RW}$ is the RW temperature, the net quark number density $n_{\rm Q}$ is a smooth function of $\theta_{\rm q}$.   
However, when $T>T_{\rm RW}$, it is discontinuous at $\theta_{\rm q}=(2k+1)\pi/3$ where $k$ is any integer, due to the degeneracy of the ground state. 
This discontinuity is called the RW transition.   
Similarly, the pressure $P$ of the system is a smooth function of $\theta_{\rm q}$ when $T<T_{\rm RW}$, but it has a cusp at $\theta_{\rm q}=(2k+1)\pi/3$ when $T>T_{\rm RW}$. 
The RW periodicity and the RW transition are confirmed by LQCD simulations
\cite{deForcrand:2002hgr,DElia:2002tig}
and $T_{\rm RW}$ is estimated as $195(\pm 1)\sim 208(\pm 5)$~MeV for the 2+1 flavor LQCD simulation~\cite{Bonati:2016pwz,Cuteri:2022vwk,Bonati:2018fvg}.

\section{Hadron resonance gas model with excluded volume effects}
\label{HRGEVE}

In this section, we review our hadron resonance gas model with EVE~\cite{Oshima:2023bip}.  
For simplicity of calculations, we assume that all baryons and antibaryons have the same volume $v_{\rm B}$. 
In numerical calculations, we put $v_{\rm B}={4\pi\over{3}}r_{\rm B}^3$ with $r_{\rm B}=0.8$~fm.  
At the baryon number chemical potential $\mu$, the net baryon number density $n_{\rm B}$ is given by 
\begin{eqnarray}
n_{\rm B}(T,\mu )&=&n_{\rm b}(T,\mu )-n_{\rm a}(T,\mu ); 
\label{EnB}
\\
n_{\rm b}(T,\mu )&=&{n_{\rm b0}(T,\mu )\over{1+v_{\rm B}n_{\rm b0}(T,\mu )}}, 
\label{Enb}
\\
n_{\rm a}(T,\mu )&=&{n_{\rm a0}(T,\mu )\over{1+v_{\rm B}n_{\rm a0}(T,\mu )}}={n_{\rm b0}(T,-\mu )\over{1+v_{\rm B}n_{\rm b0}(T,-\mu )}}, 
\nonumber\\
\label{Ena}
\end{eqnarray}
where $n_{\rm b}$ and $n_{\rm a}$ are the number density of baryons and antibaryons, respectively, and $n_{\rm b0}$ and $n_{\rm a0}$ are these quantities calculated by using the point particle approximation. 
Note that the net baryon number density $n_{\mathrm B}$ is an odd function of $\mu$, since $n_{\rm a0}(T,\mu) =n_{\rm b0}(T,-\mu )$.   
When $\mu \to \pm \infty$, 
\begin{eqnarray}
n_{\rm B}(T,\mu )\to \pm {1\over{v_{\rm B}}}. 
\label{EnB_B_close}
\end{eqnarray}
Due to EVE, the baryon number density saturates to the constant value.  

According to the thermodynamic relation, the pressure $P_B(T,\mu )$ of baryons and antibaryons is given by  
\begin{eqnarray}
P_{\rm B}(T,\mu )
&=&P_{\rm b}(T,\mu )+P_{\rm a}(T,\mu ); 
\label{PB_EVE}
\\
P_{\rm b}(T,\mu )&=&\int d\mu n_{\rm b}(T,\mu ), 
\nonumber\\
\label{Pb_EVE}
\\
P_{\rm a}(T,\mu )&=-&\int d\mu n_{\rm a}(T,\mu ) 
\nonumber\\
&=&P_{\rm b}(T,-\mu ). 
\label{Pa_EVE}
\end{eqnarray}
Note that the pressure $P_{\rm B}(T,\mu )$ is an even function of $\mu$, since $n_{\rm B}$ is the odd function of $\mu$.  
Also note that the natural boundary conditions $P_{\rm b}\to 0~(\mu \to -\infty)$ and $P_{\rm a}\to 0~(\mu \to \infty)$ are imposed.

The total pressure $P(T,\mu)$ of the system is given by
\begin{eqnarray}
P(T,\mu )=P_{\rm B}(T,\mu)+P_{\rm M}(T), 
\label{P_total}
\end{eqnarray}
where $P_{\rm M}$ is the pressure of mesons.  
For the mesons, we use an ideal Bose gas approximation. 
The other thermodynamic quantities are also calculated by using the thermodynamic relations.  
Our formalism of the HRG model with EVE is different from that of Refs.~\cite{Taradiy:2019taz,Savchuk:2019yxl,Vovchenko:2017xad} . Comparison between two models is shown in Refs.~\cite{Oshima:2023bip} and our model is consistent with the model in Ref.~\cite{Taradiy:2019taz,Savchuk:2019yxl,Vovchenko:2017xad} at least below $T_{\rm RWL}$.

In this paper, we use the Boltzmann distribution function which is a good approximation of the Fermi distribution function unless the quantum effects are not large.   
Thanks to this approximation, we can obtain simple semi analytical representations 
for thermodynamic quantities.   
The baryon number density $n_{\rm B}$ is given by 
\begin{eqnarray}
n_{\rm B}(T,\mu )&=&n_{\rm b}(T,\mu )-n_{\rm a}(T,\mu ); 
\label{EnB_BA}
\\
n_{\rm b}(T,\mu )&=&{B(T)e^{\mu /T}\over{1+v_{\rm B}B(T)e^{\m/T}}}, 
\label{Enb_BA}
\\
n_{\rm a}(T,\mu )&=&{B(T)e^{-\mu /T}\over{1+v_{\rm B}B(T)e^{-\mu /T}}}; 
\label{Ena_BA}
\\
B(T)&=&\sum_i B_i(T), 
\label{BT}
\\
B_i(T)&=&{g_{s,i}\over{2\pi^2}}\int_0^\infty dp p^2e^{-\sqrt{p^2+M_i^2}/T}, 
 \label{BiT}
\end{eqnarray}
where $M_i$ and $g_{s,i}$ are the mass and the spin degeneracy of $i$-th baryons (antibaryons), respectively.  
Note that we obtain the number density $n_{\rm b0}~(n_{\rm a0})$ of point-like baryons (antibaryons) when we put $v_{\rm B}=0$. 
The pressure $P(T,\mu )$ of baryons and antibaryons is given by  
\begin{eqnarray}
P_{\rm b}(T,\mu )&=&{T\over{v_{\rm B}}}\log{[1+v_{\rm B}B(T)e^{\mu /T}]}, 
\label{Pb_EVE_Boltz}
\\
P_{\rm a}(T,\mu )&=&{T\over{v_{\rm B}}}\log{[1+v_{\rm B}B(T)e^{-\mu /T}]}.
\label{Pa_EVE_Boltz}
\end{eqnarray}
When $B(T)=1/v_{\mathrm B}$, the baryon number density is given by 
\begin{eqnarray}
n_{\mathrm B}=B(T)\tanh{(\mu /2T)}={1\over{v_{\rm B}}}\tanh{(\mu /2T)}. 
\label{nB_close}
\end{eqnarray}
Therefore, the absolute value of $n_{\mathrm B}$ saturates to $\pm 1/v_{\rm B}$ in the limit $\mu\to \pm \infty$ as is seen in Fig.~\ref{Fig_n_TRW_mu}.  
Note that, as is seen below, the condition $B(T)=1/v_{\rm B}$ is satisfied when $T=T_{\rm RWL}$. 

\begin{figure}[t]
\centering
\centerline{\includegraphics[width=0.40\textwidth]{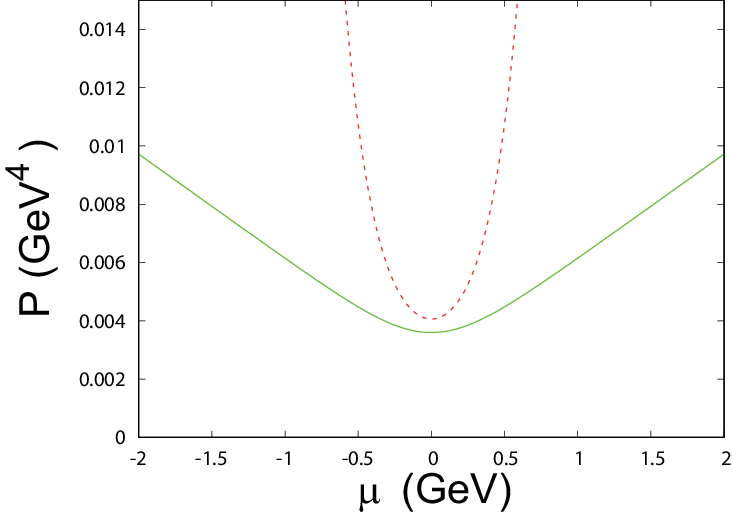}}
\caption{The $\mu$-dependence of the pressure $P$ when $T=T_{\rm RWL}=0.2103$~GeV.  
The solid and dashed lines show the results with EVE and without EVE, respectively. 
 }
 \label{Fig_P_TRW_mu}
\end{figure}

In this case, the pressure $P_{\rm B}$ is given by
\begin{eqnarray}
P_{\mathrm B}
&=&{2T\over{v_{\mathrm B}}}\log{[2\cosh{(\mu /2T)]}}, 
\label{EnB_B_r_RW_P}
\end{eqnarray}
and depends linearly on $\mu$ in the limit $\mu \to \pm\infty$ as is seen in Fig.~\ref{Fig_P_TRW_mu}. 
(Note that the meson contribution to $P$ does not depend on $\mu$. )

When $\mu$ is a pure imaginary number, 
\begin{eqnarray}
n_{\rm B}(T,\mu )
&=&{2iB(T)\sin{\theta}\over{1+\{v_{\rm B}B(T)\}^2+2v_{\rm B}B(T)\cos{(\theta)}}} ,
\label{EnB_B_i}
\\
n_{\rm b}(T,\mu )&=&{B(T)e^{i\theta}\over{1+v_{\rm B}B(T)e^{i\theta}}}, 
\label{Enb_B_i}
\\
n_{\rm a}(T,\mu )&=&{B(T)e^{-i\theta}\over{1+v_{\rm B}B(T)e^{-i\theta}}},
\label{Ena_B_i}
\end{eqnarray}
where $\mu =i\theta T$. 
If $B(T)=1/v_{\rm B}$, we obtain 
\begin{eqnarray}
n_{\rm B}
&=&iB(T)\tan{(\theta /2)}={i\over{v_{\mathrm B}}}\tan{(\theta /2)}. 
\label{EnB_B_i_RW}
\end{eqnarray}
Hence, $n_{\rm B}$ is divergent at $\theta =(2n+1)\pi$ for any integer $n$ as is seen in Fig.~\ref{Fig_n_TRW_theta}. 
This singularity is nothing but the RW-like singularity which occurs at $T=T_{\rm RWL}=0.2103$~GeV. 
The value 0.2103 GeV is slightly smaller than the value 0.2108 GeV in Ref.~\cite{Oshima:2023bip}, since we have updated the data of hadron resonances~\cite{ParticleDataGroup:2024cfk}. 
On the other hand, as was already shown in Ref.~\cite{Oshima:2023bip}, the use of the Boltzmann distribution approximation hardly changes the value of $T_{\rm RWL}$.  
If we use the nucleon gas model instead of the baryon resonances gas model, we obtain $T_{\rm RWL}=0.316$~GeV which is much higher than $T_{\rm RW}\sim 0.2$~GeV. 
The effects of the baryon resonances are very important for the realization $T_{\rm RWL}\sim T_{\rm RW}$.

\begin{figure}[t]
\centering
\centerline{\includegraphics[width=0.40\textwidth]{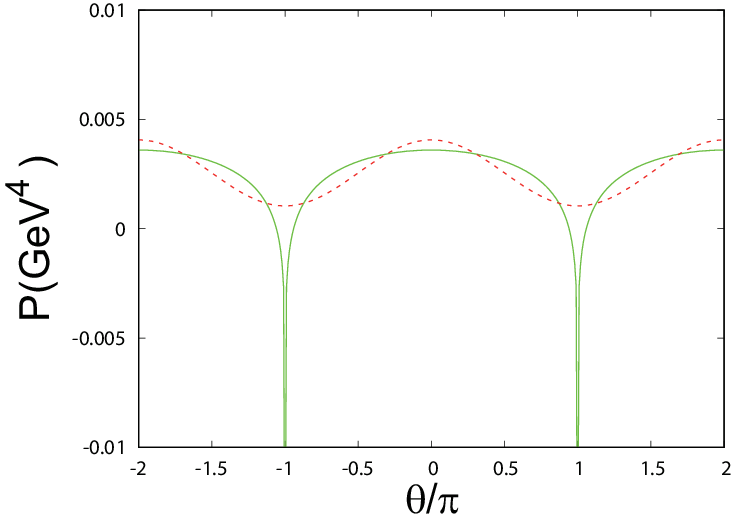}}
\caption{The $\theta$-dependence of the pressure $P$ when $T=T_{\rm RWL}=0.2103$~GeV.  
The solid and dashed lines show the results with EVE and without EVE, respectively. 
 }
 \label{Fig_P_TRW_theta}
\end{figure}

In this case, the pressure is given by
\begin{eqnarray}
P_{\mathrm B}
&=&2B(T)T\log{[2\cos{(\theta /2)]}}, 
\label{EnB_B_i_RW_P}
\end{eqnarray}
which is also divergent at  $\theta =(2n+1)\pi$ as is seen in Fig.~\ref{Fig_P_TRW_theta}

Comparing Fig.~\ref{Fig_n_TRW_theta} (Fig.~\ref{Fig_P_TRW_theta}) with Fig.~\ref{Fig_n_TRW_mu} (Fig.~\ref{Fig_P_TRW_mu}), 
we see that the saturation (linearization) of $n_{\mathrm B}$ ($P$) in the limit $\mu \to \pm\infty$ is related to the RW-like singularity at $\theta =(2n+1)\pi$. 
This indicates that the properties of the RW-like singularity in the imaginary chemical potential region contains significant information of the interaction among baryons in the large real chemical potential region.

For comparison, we show the $\theta_{\rm q}(=\theta/3)$-dependence of the number density and pressure in the PNJL model with imaginary chemical potential. The PNJL model~\cite{Fukushima:2003fw,Ratti:2005jh,Ghosh:2006qh,Megias:2004hj,Roessner:2006xn,Kashiwa:2007hw} is one of the most successful effective models of QCD and can also reproduce several important features of QCD in the region of the imaginary chemical potential when $T$ is not so small; for example, see Ref.~\cite{Kashiwa:2019ihm} as a review. 

\begin{figure}[t]
\centering
\includegraphics[width=0.39\textwidth]{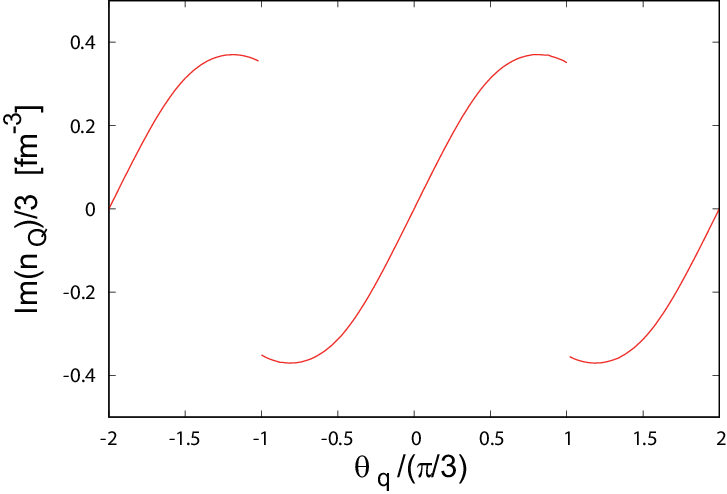}\\
\caption{The $\theta_{\rm q}$-dependence of the imaginary part of the net baryon number density $n_{\rm Q} / 3$ in the PNJL model,  when $T=T_{\rm RWL}=0.2103$~GeV; see Appendix A in Ref.~\cite{Oshima:2023bip} for the details of the model. 
Note that $T_{\rm RW}=0.201$~GeV in this model. 
}
\label{Fig_T=2103_PNJL}
\end{figure}

Fig.\,\ref{Fig_T=2103_PNJL} shows the $\theta_{\rm q}$-dependence of the net baryon number density $n_{\rm Q}/3$ obtained by the PNJL  model, when $T=T_{\rm RWL}=0.2103$~GeV.   
We see that the net baryon number density is discontinuous at $\theta =\pm \pi/3$. 
(Note that, in this calculation, $T=T_{\rm RWL}=0.2103$~GeV is larger than $T_{\rm RW}=0.201~{\rm GeV}$. 
Also note that the net baryon (quark) number density is pure imaginary in the region of the imaginary $\mu~(\mu_ q)$. )  

\begin{figure}[t]
\centering
\includegraphics[width=0.39\textwidth]{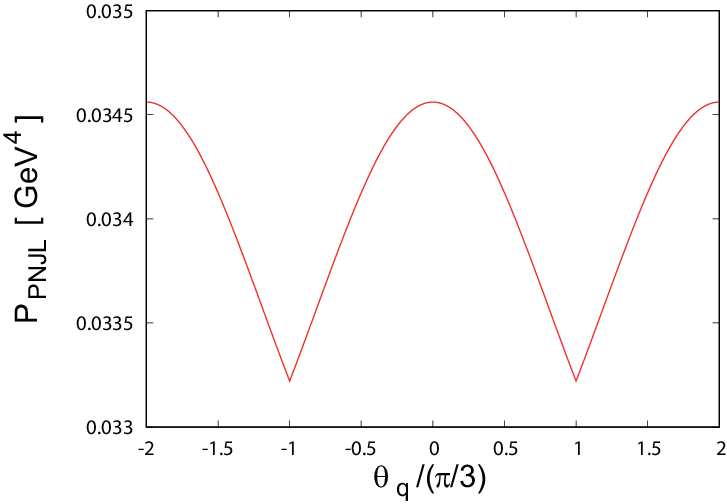}\\
\caption{The $\theta_{\rm q}$-dependence of the pressure $P_{\rm PNJL}$ in the PNJL model, when $T=T_{\rm RWL}=0.2103$~GeV. 
}
\label{Fig_PNJL_P}
\end{figure}

Fig.~\ref{Fig_PNJL_P} shows the $\theta_{\rm q}$-dependence of pressure $P_{\rm PNJL}$ in the PNJL model, when $T=T_{\rm RWL}=0.2103$~GeV. 
$P_{\rm PNJL}$ has a cusp at $\theta_{\rm q}=\pm \pi/3$.

Figure~\ref{Fig_n_TRW_theta} (Fig.~\ref{Fig_P_TRW_theta}) resembles Fig.~\ref{Fig_T=2103_PNJL} (Fig.~\ref{Fig_PNJL_P}).  
However, there are differences.  
In Fig.~\ref{Fig_n_TRW_theta} $n_{\rm B}$ has a divergence at $\theta =\pm\pi$, while,  in Fig.~\ref{Fig_T=2103_PNJL},  $n_{\rm Q}$ is discontinuous but does not have a divergence there. 
Similarly, in Fig.~\ref{Fig_P_TRW_theta} $P$ has a divergence at $\theta =\pm\pi$, while,  in Fig.~\ref{Fig_PNJL_P},  $P_{\rm PNJL}$ has a cusp but does not have divergence there. 
Therefore, this fact may indicate that $T_{\rm RWL}$ (or $T_{\rm RW}$) is a limiting temperature of baryon existence in the HRG model with EVE. 

It seems that $T_{\rm RWL}$ is slightly higher than $T_{\rm RW}$. 
Hence, our scenario of the hadron-quark transition at $\theta =(2n+1)\pi$ is as follows.  
As $T$ approaches $T_{\rm RWL}$, the baryons become unstable and transformation from baryon matter to quark matter takes place when $T$ is equal to the temperature which is slightly lower than $T_{\rm RWL}$. 
This transition temperature is nothing but the Roberge-Weiss temperature $T_{\rm RW}$. 
However, the nature of the interaction which induces the instability is not clear yet.  
Hence, in this paper, we will investigate the nature of the interaction using the thermodynamic geometry.

\section{Thermodynamic geometry}
\label{TG}

In this section, we briefly review the thermodynamic geometry according to Ref.~\cite{Ruppeiner:1995zz}.  
Consider the grand-canonical ensemble characterized by the pair of intensive variables $(\beta^1,\beta^2)=(\beta, \gamma)=(1/T, -\mu /T)$.  
In the framework of the thermodynamic geometry, we consider the 2-dimensional manifold with the coordinate $(\beta^1,\beta^2)$. 
The metric tensor in the manifold is given by 
\begin{eqnarray}
g_{i,j}=\phi_{\beta^i,\beta^j}={\partial^2\phi \over{\partial \beta^i \partial \beta^j}}~~~(i,j=1,2);~~~\phi ={P\over{T}}, 
\nonumber\\
\label{metric}
\end{eqnarray}
where $P$ is the pressure of the system. 

The probability of a fluctuation from the state $(\beta,\gamma )$ to $(\beta +\delta \beta ,\gamma +\delta \gamma)$ is proportional to 
\begin{eqnarray}
\sqrt{g}\exp{\left( -{dl^2\over{2}}\right)}, 
\label{fluct}
\end{eqnarray}
where 
\begin{eqnarray}
g={\rm det}
\left[ 
\begin{array}{cc}
g_{11} & g_{12} \\ 
g_{21} & g_{22}
\end{array}
\right]
=
{\rm det}
\left[
\begin{array}{cc}
\phi_{\beta\beta} & \phi_{\beta\gamma} \\ 
\phi_{\gamma\beta} & \phi_{\gamma\gamma}
\end{array}
\right]. 
\label{detg}
\end{eqnarray}
with the symmetric condition $g_{ij}=g_{ji}~(i,j=1,2)$ and 
\begin{eqnarray}
dl^2&=g_{11}d\beta^2 +2g_{12}d\beta d\gamma 
+g_{22}d\gamma^2
\nonumber\\
&=(d\beta, d\gamma )
\left(
\begin{array}{cc}
g_{11} & g_{12} \\
g_{21} & g_{22} 
\end{array}
\right)
\left(
\begin{array}{c}
d\beta \\
d\gamma 
\end{array}
\right)
\nonumber\\
&=g_{11}\left( d\beta +{g_{12}\over{g_{11}}}d\gamma\right )^2 
+{g\over{g_{11}}}d\gamma^2
\nonumber\\
\label{dl2}
\end{eqnarray}
is the line element. 
The condition that $g_{11}>0$ and $g>0$ is required for the thermodynamic stability. 
In this case, $dl^2>0$ unless $d\beta=d\gamma=0$. 
When $g<0$, the system is thermodynamically unstable. 

The scalar curvature, which is called the thermodynamic curvature in the framework of the thermodynamic geometry, is given by 
\begin{eqnarray}
R=\frac{h}{2g^2};~~~
h={\rm det}\left[
\begin{array}{ccc}
\phi_{\beta\beta} & \phi_{\beta\gamma} & \phi_{\gamma\gamma} \\ 
\phi_{\beta\beta\beta} & \phi_{\beta\beta\gamma} & \phi_{\beta\gamma\gamma} \\ 
\phi_{\beta\beta\gamma} & \phi_{\beta\gamma\gamma} & \phi_{\gamma\gamma\gamma}
\end{array}
\right], 
\label{R}
\end{eqnarray}
where
\begin{eqnarray}
\phi_{\beta\beta\beta}={\partial^3 \phi\over{\partial \beta^3}},~~~ 
\phi_{\beta\beta\gamma}={\partial^3 \phi\over{\partial \beta^2\partial\gamma}}, 
\nonumber\\
\phi_{\beta\gamma\gamma}={\partial^3 \phi\over{\partial\beta\partial \gamma^2}}, ~~~
\phi_{\gamma\gamma\gamma}={\partial^3 \phi\over{\partial \gamma^3}}. 
\label{phi_d3}
\end{eqnarray}

The scalar curvature $R$ can be used to investigate the nature of the interaction. 
It is expected that the interaction in the system is repulsive (attractive), when $R>0$ ($R<0$).  
Using this criterion, it can be concluded that the ideal fermion gas has the repulsive nature, while the ideal boson gas exhibits an attractive one~\cite{Janyszek:1990wdh}.   
Hence, the interaction in baryon (meson) gas is expected to be repulsive (attractive).

The thermodynamic geometry can be also used to determine the phase diagram. 
In $\mu$-$T$ plane, the curve on which $R=0$ may be the transition line of the phase transition. 
Castorina, Imbrosciano and Lanteri used the thermodynamic geometry to investigate the QCD phase diagram in $\mu$-$T$ plane~\cite{Castorina:2018ayy}. 
Since there is a sign problem in finite density LQCD, they used the power series expansion method to estimate $\phi$ and calculated the scalar curvature $R$. They used the $R=0$ criterion to determine the pseudo-critical temperature of the crossover transition. 
The obtained pseudo-critical temperature is consistent with that obtained by using the standard criterion in the LQCD.  
They also used the HRG model to estimate $\phi$ and calculated $R$. They found that the pseudo-critical temperature obtained in the HRG model is somewhat higher than that in LQCD. 
In Ref.~\cite{Castorina:2018gsx}, they performed the similar analyses but used also the HRG model with EVE. 
In this case, they found that the pseudo-critical temperature obtained by the HRG model with EVE is consistent with that in LQCD. 

The scalar curvature $R$ has the same dimension as the spatial volume. 
Hence, it is expected that, in proximity of a second order phase transition, $R$ is proportional to $\xi^3$ where $\xi$ is the correlation length and diverges when the second order phase transition takes place. 
Castorina, Lanteri and Mancani ~\cite{Castorina:2019jzw} studied the thermodynamic geometry in Nambu--Jona-Lasinio model. 
They found that the scalar curvature $R$ shows the divergent behavior at the critical temperature of the chiral phase transition when the current quark mass is zero. 
They also found that $R$ becomes large when the system approaches the critical point when the current quark mass is finite. 
Zhang, Wan and Ruggieri did the similar studies by using the Quark-meson model~\cite{Zhang:2019neb}. 
Castorina, Lanteri and Ruggieri studied the fluctuations and the thermodynamic geometry of the chiral phase transition in the Quark-meson model~\cite{Castorina:2020vbh}.  
Murgana, Greco, Ruggieri and Zappal\`a studied the thermodynamic geometry of the Quark-meson model using the functional renormalization group~\cite{Murgana:2023pyx}.  

At a glance, the two criteria, namely, the $R=0$ criterion and the $|R|=\infty$ criterion, contradict each other.  
However, in many cases, the divergence of $R$ is accompanied by the change of sign of $R$~\cite{Murgana:2023pyx}.    
In particular, the sign change of $R$ takes place in the vicinity of the critical point where $R$ shows the divergent behavior~\cite{Murgana:2023pyx}.  
As will be seen later, there is no divergence of $R$ in the HRG model when the chemical potential $\mu$ is real. 
Hence, we mainly use the $R=0$ criterion to investigate the phase structure in the HRG model as was in the previous studies~\cite{Castorina:2018ayy, Castorina:2018gsx}.

When $\mu$ is pure imaginary, $d\gamma^2 \le 0$.  
Hence, in this case, we should use the variable $({\beta^\prime}^1,{\beta^\prime}^2)=(\beta, \theta)=(1/T, -i\mu /T)$ instead of 
$(\beta^1,\beta^2)=(\beta, \gamma)=(1/T, -\mu /T)$.   
The metric tensor in the manifold is given by 
\begin{eqnarray}
g_{i,j}^\prime =\phi_{{\beta^\prime}^i,{\beta^\prime}^j}={\partial^2\phi \over{\partial {\beta^\prime}^i \partial {\beta^\prime}^j}}~~~(i,j=1,2). 
\nonumber\\
\label{metric_im}
\end{eqnarray}
The line element is given by 
\begin{eqnarray}
d{l^\prime}^2&=g_{11}^\prime d\beta^2 +2g_{12}^\prime d\beta d\theta
+g_{22}^\prime d\theta^2
\nonumber\\
&=g_{11}^\prime \left( d\beta +{g_{12}^\prime \over{g_{11}^\prime }}d\theta \right )^2 
+{g^\prime\over{g_{11}^\prime }}d\theta^2, 
\nonumber\\
\label{dl2_2}
\end{eqnarray}
where 
\begin{eqnarray}
g^\prime ={\rm det}
\left[ 
\begin{array}{cc}
g_{11}^\prime  & g_{12}^\prime \\ 
g_{21}^\prime & g_{22}^\prime
\end{array}
\right]
=
{\rm det}\left[
\begin{array}{cc}
\phi_{\beta\beta} & \phi_{\beta\theta} \\ 
\phi_{\theta\beta} & \phi_{\theta\theta}
\end{array}
\right]=-g. 
\label{g_2}
\end{eqnarray}
The condition that $g_{11}^\prime=g_{11}>0$ and $g^\prime =-g>0$ is required for the thermodynamic stability.

The scalar curvature is given by
\begin{eqnarray}
R^\prime =\frac{h^\prime }{2{g^\prime}^2}=R;~~~~~
h^\prime 
=
{\rm det}\left[
\begin{array}{ccc}
\phi_{\beta\beta} & \phi_{\beta\theta} & \phi_{\theta\theta} \\
\phi_{\beta\beta\beta} & \phi_{\beta\beta\theta} & \phi_{\beta\theta\theta} \\
\phi_{\beta\beta\theta} & \phi_{\beta\theta\theta} & \phi_{\theta\theta\theta}
\end{array}
\right]=h,  
\nonumber\\
\label{R_2}
\end{eqnarray}
where 
\begin{eqnarray} 
\phi_{\beta\beta\theta}={\partial^3 \phi\over{\partial \beta^2\partial\theta}},~~
\phi_{\beta\theta\theta}={\partial^3 \phi\over{\partial\beta\partial \theta^2}},~~
\phi_{\theta\theta\theta}={\partial^3 \phi\over{\partial \theta^3}}. 
\label{phi_d3_im}
\end{eqnarray}

As is seen in the next section, the stability condition is not satisfied in the wide region, when $\mu$ is imaginary. 
Nevertheless we can consider the scalar curvature $R$ at imaginary $\mu$ as the analytic continuation from the one at real $\mu$. 
Inversely, we can estimate $R$ at real $\mu$ as the analytical continuation from the one in the imaginary chemical potential region where there is no sign problem in LQCD calculation.

\section{Scalar curvature}
\label{Scurve}

In this section, we investigate the scalar curvature $R$ in the HRG models with and without EVE. 
Since the calculation of $R$ is difficult at low temperature, we consider the cases only in the region of $T=0.025-0.3$ GeV.   
Note that, since $\gamma$ depends not only on $\mu$ but also on $T$, the relation between two coordinates $(T,\mu)$ and $(\beta,\gamma )$ is not simple and we obtain
\begin{eqnarray}
{\partial\over{\partial\beta}}&=&-T^2{\partial\over{\partial T}}-T\mu{\partial\over{\partial \mu}},~~~~~
{\partial\over{\partial\gamma}}=-T{\partial\over{\partial \mu}}. 
\label{coordinate}
\end{eqnarray}

 \subsection{Results at $\mu =0$. }

\begin{figure}[t]
\centering
\centerline{\includegraphics[width=0.40\textwidth]{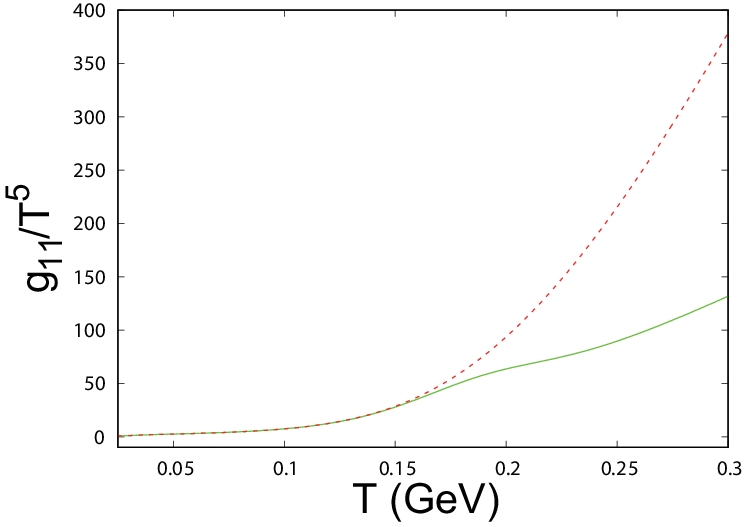}}
\caption{The $T$-dependence of $g_{11}$ when $\mu =0$.  
The solid and dashed lines show the results in the HRG model with EVE and without EVE, respectively. 
 }
\label{Fig_G11_T_mu=0}
\end{figure}

Figure~\ref{Fig_G11_T_mu=0} shows the $T$-dependence of $g_{11}~(=\phi_{\beta\beta})$ when $\mu =0$.  
$g_{11}$ is always positive. 

\begin{figure}[t]
\centering
\centerline{\includegraphics[width=0.40\textwidth]{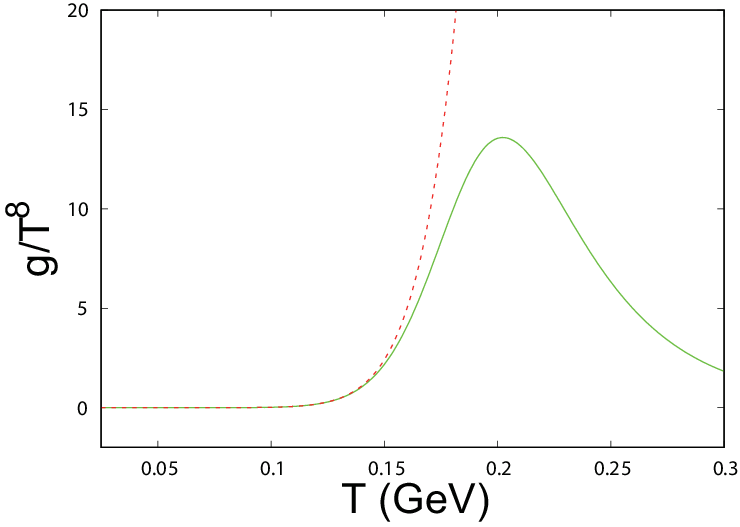}}
\caption{The $T$-dependence of the determinant $g$ of the metric when $\mu =0$.  
The solid and dashed lines show the results in the HRG model with EVE and without EVE, respectively. 
 }
 \label{Fig_G_T_mu=0}
\end{figure}

Figure~\ref{Fig_G_T_mu=0} shows the $T$-dependence of the determinant $g$ of the metric when $\mu =0$.  
As well as $g_{11}$, $g$ is also always positive. 
Hence, the line element (\ref{dl2}) is positive definite and the system is thermodynamically stable. 

\begin{figure}[t]
\centering
\centerline{\includegraphics[width=0.40\textwidth]{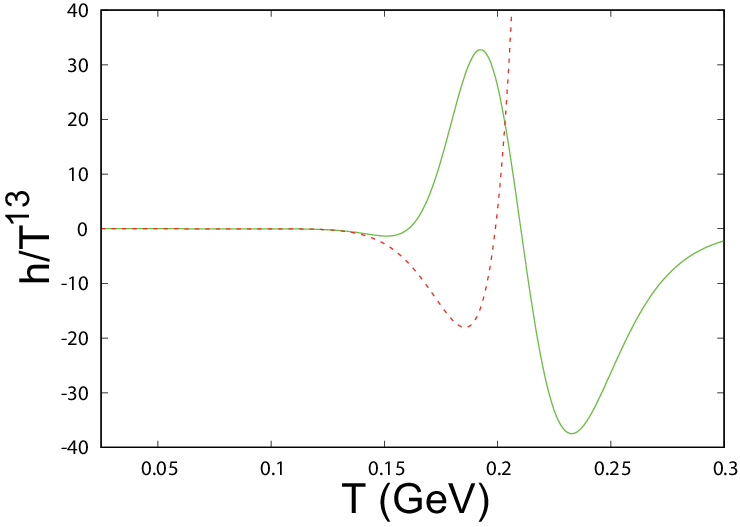}}
\caption{The $T$-dependence of the numerator $h$ of the scalar curvature when $\mu =0$. 
The solid and dashed lines show the results in the HRG model with EVE and without EVE, respectively. 
 }
 \label{Fig_H_T_mu=0}
\end{figure}

Figure~\ref{Fig_H_T_mu=0} shows the $T$-dependence of the numerator $h$ of the scalar curvature $R$ when $\mu =0$.  
When EVE is absent, $h<0$ for $T<0.199$ GeV and $h>0$ for $T>$0.199 GeV. 
When EVE is present, $h<0$ for $T<$0.161 GeV or $T>$0.2103 GeV, while $h>0$ for 0.161 GeV$<T<$0.2103 GeV.  

\begin{figure}[t]
\centering
\centerline{\includegraphics[width=0.40\textwidth]{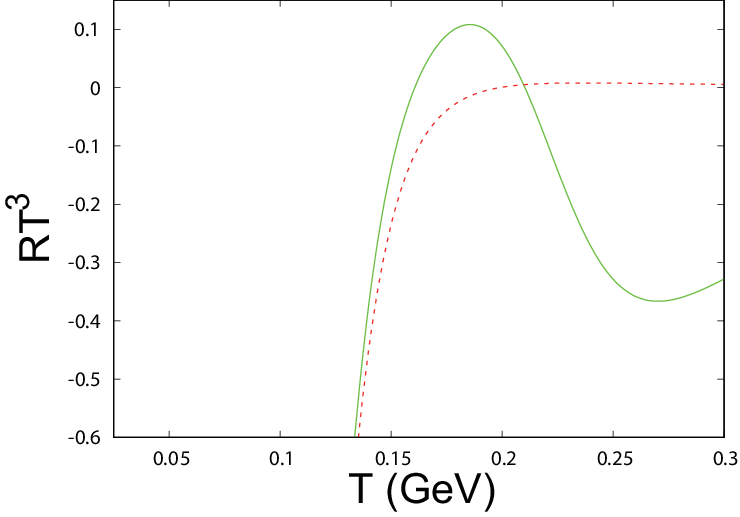}}
\caption{The $T$-dependence of the scalar curvature $R$ when $\mu =0$.  
The solid and dashed lines show the results in the HRG model with EVE and without EVE, respectively. 
 }
 \label{Fig_R_T_mu=0}
\end{figure}

Figure~\ref{Fig_R_T_mu=0} shows the $T$-dependence of the scalar curvature $R$ when $\mu =0$. 
When EVE is absent, $R<0$ for $T<$0.199 GeV and $R>0$ for $T>$0.199 GeV. 
When EVE is present, $R<0$ for $T<$ 0.161 GeV or $T>$0.2103 GeV, while $R>0$ for 0.161 GeV$<T<$0.2103 GeV. 
The sign of $R$ is the same as the one of $h$, since $g^2>0$.   

At $\mu =0$, the $n$-th derivative of $\phi$ with respect to $\gamma$ vanishes when $n$ is odd. 
Then, $h$ is reduced to 
\begin{eqnarray}
h&=&h_1h_2; 
\nonumber\\
h_1&=&\phi_{\beta\gamma\gamma}={\partial \phi_{\gamma\gamma}\over{\partial \beta}},
\nonumber\\
h_2&=&\phi_{\gamma\gamma}\phi_{\beta\beta\beta}-\phi_{\beta\beta}\phi_{\beta\gamma\gamma}.
\label{H_simple}
\end{eqnarray}
Hence, $h=0$ means $h_1=0$ and/or $h_2=0$.  
Note that $h_2$ can be rewritten as 
\begin{eqnarray}
h_2&=&-\phi_{\beta\beta}^2{\partial\over{\partial\beta}}  {\phi_{\gamma\gamma}\over{\phi_{\beta\beta}}} ,  
\label{H_simple_2}
\end{eqnarray}
when $g_{11}=\phi_{\beta\beta}\neq 0$. 
As will be seen in the next section, $\phi_{\gamma\gamma}$ ($\phi_{\gamma\gamma}/\phi_{\beta\beta}$) has its maximum value at $T=T_1=0.2103$~GeV ($T=T_2=0.161$ GeV) in the HRG model with EVE. 

\begin{figure}[t]
\centering
\centerline{\includegraphics[width=0.40\textwidth]{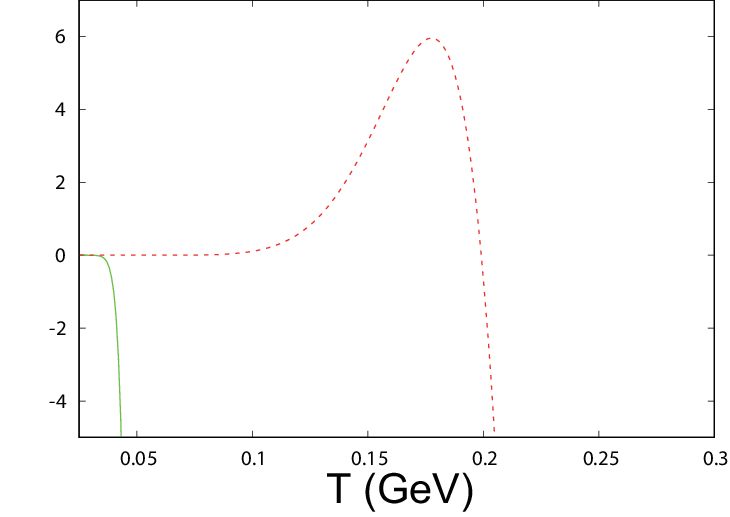}}
\caption{The $T$-dependence of $\phi_{\beta\gamma\gamma}/T^{4}$(the solid line) and $(\phi_{\gamma\gamma}\phi_{\beta\beta\beta}-\phi_{\beta\beta}\phi_{\beta\gamma\gamma})/T^{9}$ (the dashed line) in the HRG model without EVE when $\mu =0$. 
The result of  $\phi_{\beta\gamma\gamma}/T^4$ is multiplied by the factor $10^7$.   
 }
 \label{Fig_hzerocT_mu=0_pt}
\end{figure}

Figure~\ref{Fig_hzerocT_mu=0_pt} shows $h_1$ and $h_2$ in the HRG model without EVE. 
$h_1$ is always negative, while 
$h_2$ is positive for $T<0.199$ GeV and negative for $T>0.199$ GeV. 
The sign of $h$ is the opposite to that of $h_2$, since $h_1<0$.  

\begin{figure}[t]
\centering
\centerline{\includegraphics[width=0.40\textwidth]{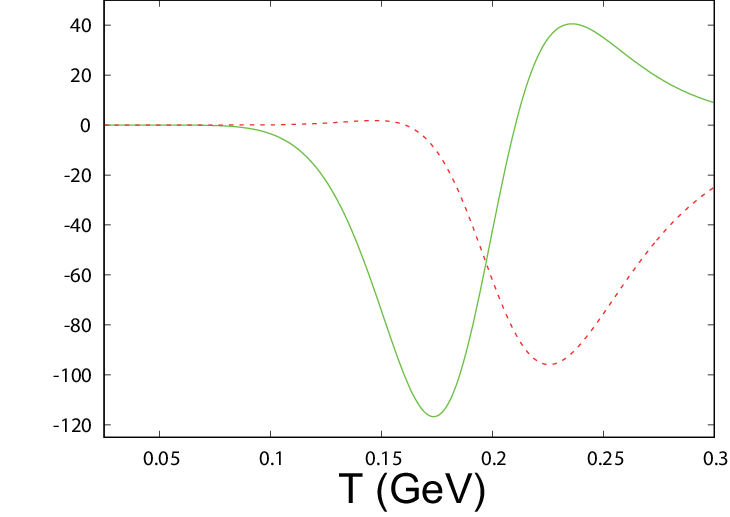}}
\caption{The $T$-dependence of $\phi_{\beta\gamma\gamma}/T^4$(the solid line) and $(\phi_{\gamma\gamma}\phi_{\beta\beta\beta}-\phi_{\beta\beta}\phi_{\beta\gamma\gamma})/T^9$ (the dashed line) in the HRG model with EVE when $\mu =0$. 
The result of  $\phi_{\beta\gamma\gamma}/T^4$ is multiplied by the factor 100.  
 }
 \label{Fig_hzerocT_mu=0_EVE}
\end{figure}

Figure~\ref{Fig_hzerocT_mu=0_EVE} shows $h_1$ and $h_2$ in the HRG model with EVE. 
$h_1$ is negative for $T<0.2103$~GeV and positive for $T>0.2103$~GeV.  
$h_2$ is positive for $T<0.161$ GeV and negative for $T>0.161$~GeV.  
In the case of the HRG model with EVE, direct calculations show 
\begin{eqnarray}
\phi_{\beta\gamma\gamma}=-{2T^2B^\prime (T)(1-vB(T))\over{(1+vB(T))^3}}; ~~~~~
B^\prime (T)\equiv{dB(T)\over{dT}}
\nonumber\\
\label{Eq_mu=0_zero}
\end{eqnarray}
at $\mu =0$. 
Hence, $h_1=\phi_{\beta\gamma\gamma}=0$, $h=0$ and $R=0$ when $B(T)=1/v_{\rm B}$. 
The condition $B(T)=1/v_{\rm B}$ is nothing but the condition when the singularity occurs at $\theta =\pi$. 
Therefore, $R=0$ when $\mu =0$ and $T=T_1=T_{\rm RWL}=0.2103$~GeV. 
We see that $T_{\rm RWL}$ is a special temperature even at $\mu =0$. 

\subsection{Results at finite real $\mu$. }

We confirmed that $g_{11}$ and $g$ are always positive as in the case at $\mu =0$ when $\mu$ is finite and real. 
Hence, the line element (\ref{dl2}) is positive definite in this case and the system is thermodynamically stable.

\begin{figure}[t]
\centering
\centerline{\includegraphics[width=0.40\textwidth]{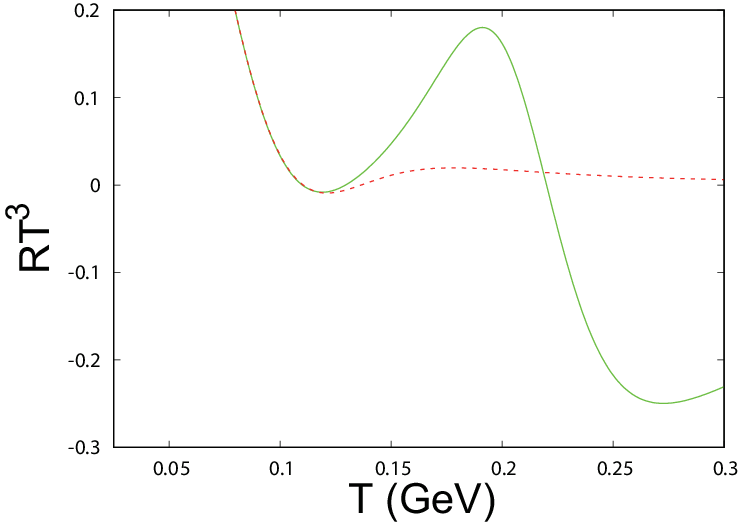}}
\caption{The $T$-dependence of the scalar curvature $R$ when $\mu =0.2$~GeV.   
The solid and dashed lines show the results in the HRG model with EVE and without EVE, respectively. 
 }
 \label{Fig_R_T_mu=200}
\end{figure}

Figure~\ref{Fig_R_T_mu=200} shows the $T$-dependence of the scalar curvature $R$ when $\mu =0.2$~GeV.  
When EVE is absent, $R<0$ for $0.110~{\rm GeV} <T<0.137$ GeV and $R>0$ for $T<0.110$ GeV or $T>0.137$ GeV. 
When EVE is present, $R<0$ for 0.110~GeV$<T<$0.129 GeV or $T>0.220$~GeV, while $R>0$ for $T<0.110$ GeV and 0.129 GeV$<T<$0.220 GeV. 

\begin{figure}[t]
\centering
\centerline{\includegraphics[width=0.40\textwidth]{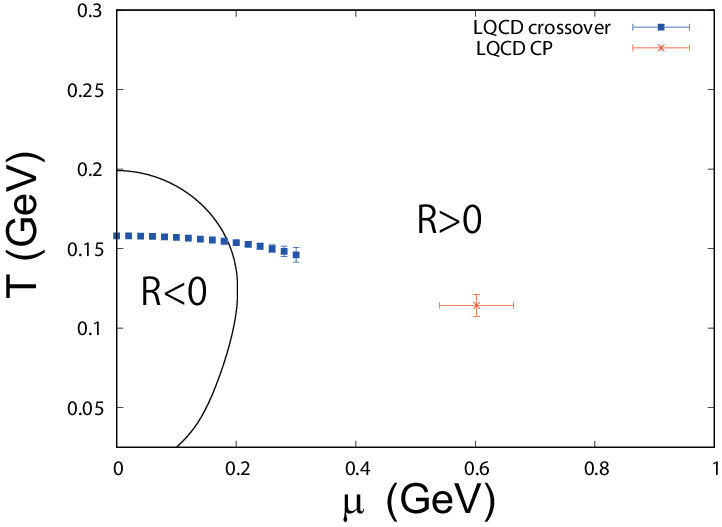}}
\caption{The solid line represents the curve determined by the condition $R=0$ on the $\mu$-$T$ plane in the HRG model without EVE. 
The squares with the error bar show the LQCD crossover line taken from Ref. ~\cite{Borsanyi:2020fev}. 
The cross with error bars shows the LQCD predicted critical point taken from Ref.~\cite{Shah:2024img}. 
}
 \label{Fig_PD}
\end{figure}

Figure~\ref{Fig_PD} shows the curve on which $R=0$ is satisfied on the $\mu$-$T$ plane in the HRG model without EVE. 
When $T$ is large and/or $\mu$ is large, $R$ is positive. 
At small $\mu $, the pseudo-critical temperature obtained by the $R=0$ criterion is much larger than the LQCD crossover temperature. 
This result is consistent with the previous studies~\cite{Castorina:2018ayy,Castorina:2018gsx}. 
There is no phase structure in the large $\mu$ region.

\begin{figure}[t]
\centering
\centerline{\includegraphics[width=0.40\textwidth]{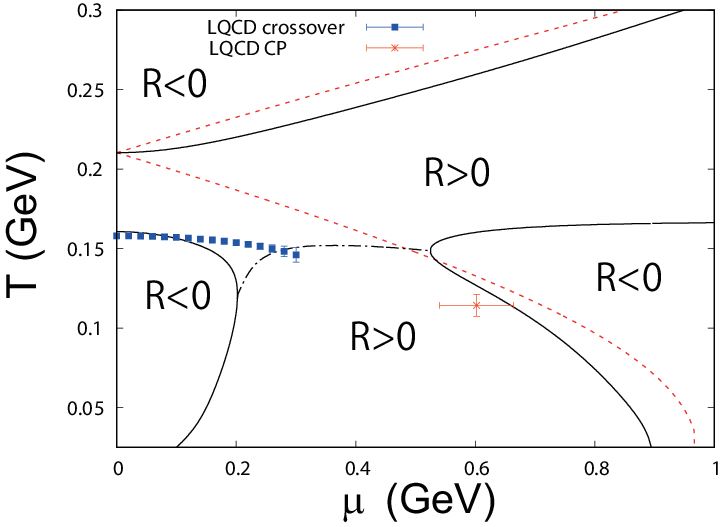}}
\caption{The solid lines represent the curves determined by the condition $R=0$ on the $\mu$-$T$ plane in the HRG model with EVE.  
The dash-dotted line represents the curve on which $|RT^3|$ has its local minimum for the fixed value of $\mu$ in the region where 0.202~GeV$<\mu <0.525$~GeV and $T<T_{\rm RWL}$. 
The lower (upper) dashed line represents the curve determined by the condition $n_{\rm b}=1/(2v_{\rm B})$ ($n_{\rm a}=1/(2v_{\rm B}$)). 
The meaning of the symbols is the same as in Fig.~\ref{Fig_PD}. 
 }
 \label{Fig_PD_EVE}
\end{figure}

Figure~\ref{Fig_PD_EVE} is the same as Fig.~\ref{Fig_PD} but the HRG model with EVE is used.  
In this case, $R<0$ when $T$ is very high and when $T$ is small and $\mu$ is small or very large. 
In the small $\mu$ region, the lower curve of the $R=0$ criterion is consistent with the LQCD crossover temperature up to $\mu \sim 0.1$~GeV. 
This result is consistent with the previous study~\cite{Castorina:2018gsx}. 
The crossover temperature line meets the $|RT^3|$ local minimum curve at $\mu \sim 0.25$~GeV. 
We have also calculated $R$ in the large $\mu$ region. 
The critical point (CP) predicted by the LQCD calculation lies just below the right-lower curve of the $R=0$ criterion. 
This small disagreement indicates that the HRG model with EVE is not applicable beyond the critical point. 
In the next section, we investigate the limitation of the HRG model with EVE.

In Fig.~\ref{Fig_PD_EVE}, the curve of the condition  $n_{\rm b}=1/(2v_{\rm B})$ ($n_{\rm a}=1/(2v_{\rm B})$) is also shown. 
This condition can be rewritten as $n_{\rm b0}=1/v_{\rm B}$ ($n_{\rm a0}=1/v_{\rm B}$). 
These curves seem to relate with the curves of the $R=0$ criterion. 
In the next section, we investigate the physical meaning of these curves.  

\subsection{Results at imaginary $\mu$. }

\begin{figure}[t]
\centering
\centerline{\includegraphics[width=0.40\textwidth]{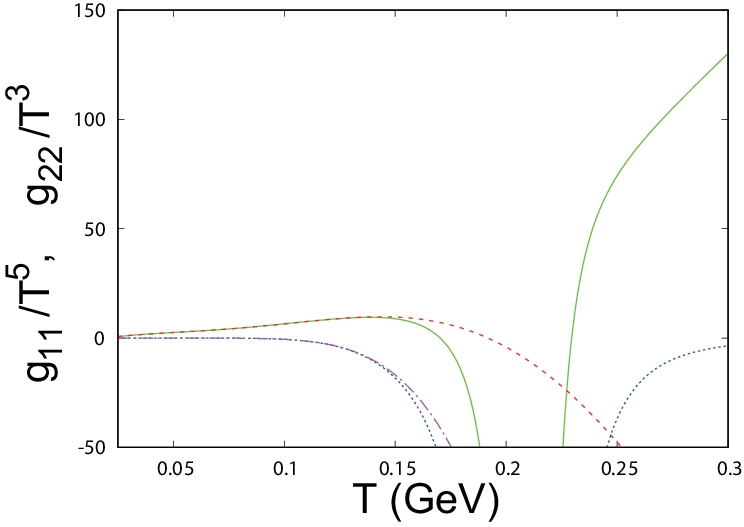}}
\caption{The solid (dashed) line shows the $T$-dependence of the $g_{11}$ in the HRG model with (without) EVE when $\theta=\pi$.  
The dotted (dash-dotted) line shows the $T$-dependence of the $g_{22}$ in the HRG model with (without) EVE when $\theta=\pi$.  
$g_{22}/T^3$ is multiplied by the factor 200. 
 }
\label{Fig_G11_G22_T_theta=pi}
\end{figure}

Figure~\ref{Fig_G11_G22_T_theta=pi} shows the $T$-dependence of $g_{11}$ and $g_{22}$ when $\theta=\pi$. 
In the HRG model without EVE,  $g_{11}>0$ for $T<0.192$~GeV and $g_{11}<0$ for $T>0.192$~GeV, while  $g_{22}$ is always negative. 
In the HRG model with EVE,  $g_{11}<0$ for 0.170~GeV$<T<$0.229~GeV and $g_{11}>0$ elsewhere, while  $g_{22}$ is always negative.  
As well as in the case of $\mu =0$, the $n$-th derivative with respect to $\gamma$ ($\theta$) vanishes at $\theta =\pi$ when $n$ is odd. 
Hence, $g$ is reduced to $g=g_{11}g_{22}$
Therefore, the sign of $g$ is opposite to that of $g_{11}$, since $g_{22}$ is always negative. 
$g=0$ at $T=0.192$~GeV ($T=0.170,0.229$~GeV) in the HRG model without EVE (with EVE), since $g_{11}=0$ there. 

\begin{figure}[t]
\centering
\centerline{\includegraphics[width=0.40\textwidth]{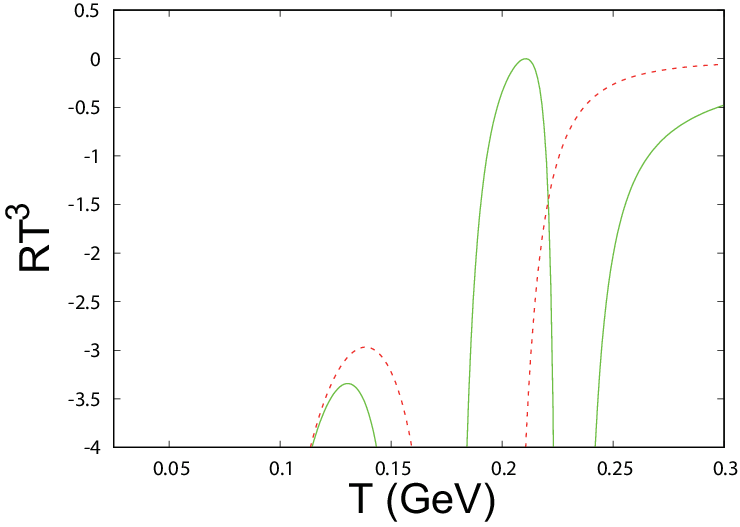}}
\caption{The $T$-dependence of the scalar curvature $R$ when $\theta =\pi$.    
The solid and dashed lines show the results in the HRG model with EVE and without EVE , respectively. 
 }
 \label{Fig_R_T_theta=pi}
\end{figure}

Figure~\ref{Fig_R_T_theta=pi} shows the $T$-dependence of the scalar curvature $R$ when $\theta=\pi$.  
In the HRG model without EVE, $R$ is always negative but diverges to $-\infty$ at $T=0.192$~GeV since $g=0$ there.  
When $\theta =\pi$, the number density of the point-like baryon is given by 
$n_{b0}=-B(T)$. 
Therefore, the baryon contribution to $\phi$ is negative and cancels the positive meson contribution. 
This cancellation yields the zeros of $g_{11}$ and $g$.    
In the HRG model with EVE, $R$ is positive for $T_{\rm RWL}<T<$0.211~GeV and negative elsewhere, and diverges to $-\infty$ at $T=0.170$~GeV and $T=0.229$~GeV where $g=0$.   
At $T=T_{\rm RWL}$, $g$ and $h$ diverge and $R$ cannot be defined. 
However, the divergent factor $(1-vB(T))^{-8}$ in $h$ is canceled by the same factor in $g^2$. 
Hence, we can calculate $R$ in the limit $T\to T_{\rm RWL}$. 
Analytical calculation shows  $R\to 0$ when $T\to T_{\rm RWL}$. 
In Fig.~\ref{Fig_R_T_theta=pi}, we use this value of $R$ for $T=T_{\rm RWL}$. 
Singular behaviors of the thermodynamical quantities at $T_{\rm RWL}$ are related with the change of the sign of $R$.  

\begin{figure}[t]
\centering
\centerline{\includegraphics[width=0.40\textwidth]{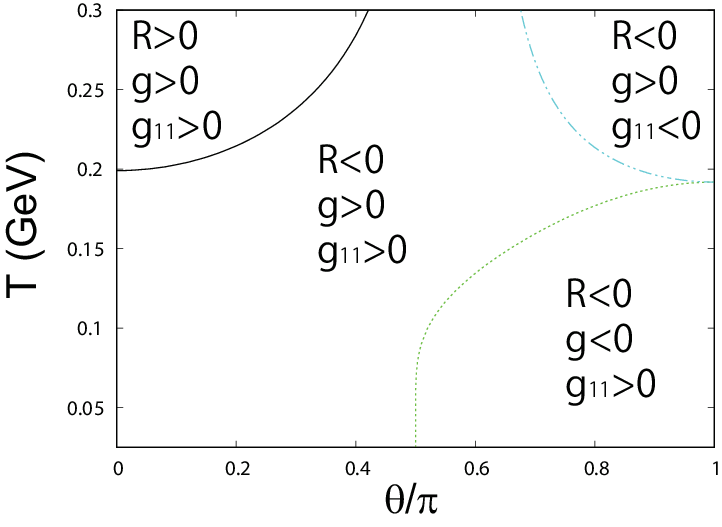}}
\caption{The solid line represents the curve determined by the condition $R=0$ on the $\theta$-$T$ plane in the HRG model without EVE.  
The dotted (dash-dot-dotted) line represents the curve determined by the condition $g=0$ ($g_{11}=0$). }
 \label{Fig_PD_theta-T}
\end{figure}

Figure~\ref{Fig_PD_theta-T} shows the curve determined by the condition $R=0$ on the $\theta$-$T$ plane in the HRG model without EVE. 
$R$ is positive in the top left corner of the graph and negative elsewhere. 
In the bottom right corner of the graph, $R<0$, $g_{11}>0$ and $g<0$. 
In this region, the thermodynamic stability condition is formally satisfied since $g^\prime_{11}=g_{11}$ and $g^\prime =-g>0$. 
(However, in the imaginary $\mu$ region, thermodynamic interpretation itself may be difficult since $n_{\rm B}$ is pure imaginary. )
The system is thermodynamically unstable elsewhere. 
In the top right corner, there is a region where $g_{11}$ is negative  

\begin{figure}[t]
\centering
\centerline{\includegraphics[width=0.42\textwidth]{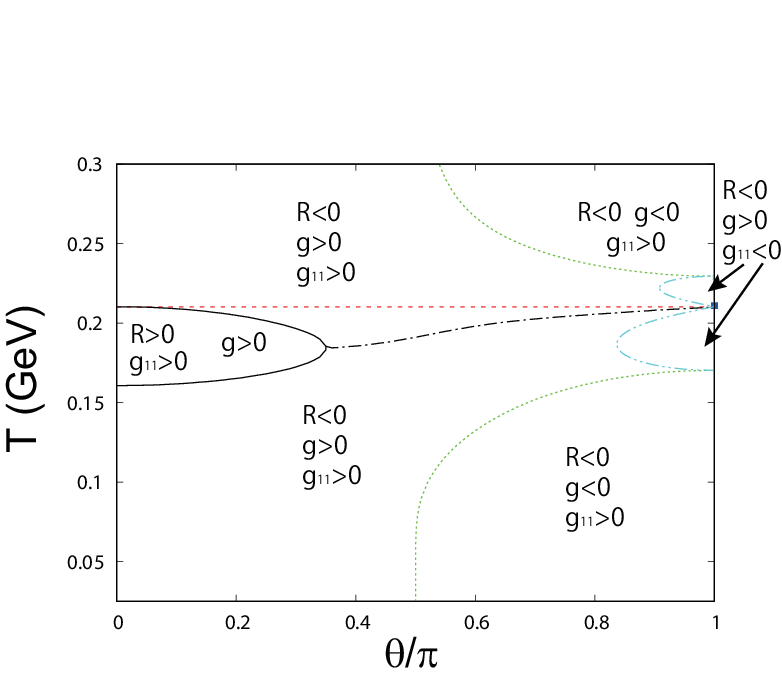}}
\caption{The solid line represents the curve determined by the condition $R=0$ on the $\theta$-$T$ plane in the HRG model with EVE.  
The dash-dotted line represents the curve on which $|RT^3|$ has its minimum for the fixed value of $\theta$ in the region of $0.35<\theta <0.9914$ where the zero point of $R$ does not exist.  
The square is the RWL point $(\theta, T)=(\pi,T_{\rm RWL})$. 
The dotted (dash-dot-dotted) line represents the curve determined by the condition $g=0$ ($g_{11}=0$). 
The dashed line represents the curve determined by the condition $|n_{\rm b0}|=B(T)=1/v_{\rm B}$. Note that $|n_{\rm a0}|=|n_{\rm b0}|$ when $\mu$ is imaginary. 
 }
 \label{Fig_PD_theta-T_EVE}
\end{figure}

\begin{figure}[t]
\centering
\centerline{\includegraphics[width=0.40\textwidth]{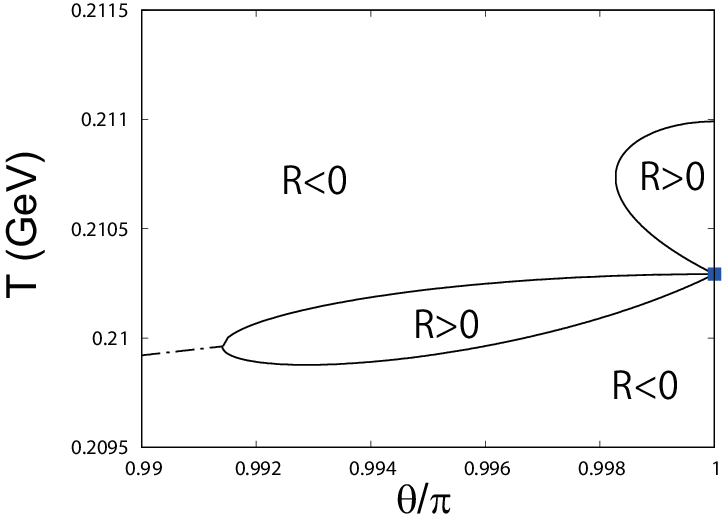}}
\caption{The solid line represents the curve determined by the condition $R=0$ on the $\theta$-$T$ plane in the HRG model with EVE.  
The dash-dotted line represents the curve on which $|RT^3|$ has its minimum for the fixed value of $\theta$ in the region of $0.99\leq \theta <0.9914$ where the zero point of $R$ does not exists. 
The square is the RWL point $(\theta, T)=(\pi,T_{\rm RWL})$. 
 }
 \label{Fig_PD_theta-T_EVE_fs}
\end{figure}

Figure~\ref{Fig_PD_theta-T_EVE} is the same as Fig.~\ref{Fig_PD_theta-T} but the HRG model with EVE is used.  
In this case, $R>0$ in the region of intermediate temperature and $\theta <0.36\pi$, and in the vicinity of the RWL point $(\theta, T)=(\pi, T_{\rm RWL})$. 
The fine structure of the curve of $R=0$ in the vicinity of the RWL point is shown in Fig.~\ref{Fig_PD_theta-T_EVE_fs}.  
$R$ is negative elsewhere. 
The curve of $R=0$ near the $T$ axis is connected to the RWL region by the curve of $|RT^3|_{\rm min}$. 
There are two regions where $g^\prime =-g>0$ and $g_{11}^\prime =g_{11}>0$. 
There are also two regions where $g_{11}$ is negative. 
The curve determined by the condition $|n_{b0}|=B(T)=1/v_{\rm B}$ is the horizontal line which connects the upper zero point of $R$ at $\theta =0$ with the RWL point at $\theta =\pi$. 

\begin{figure}[t]
\centering
\centerline{\includegraphics[width=0.40\textwidth]{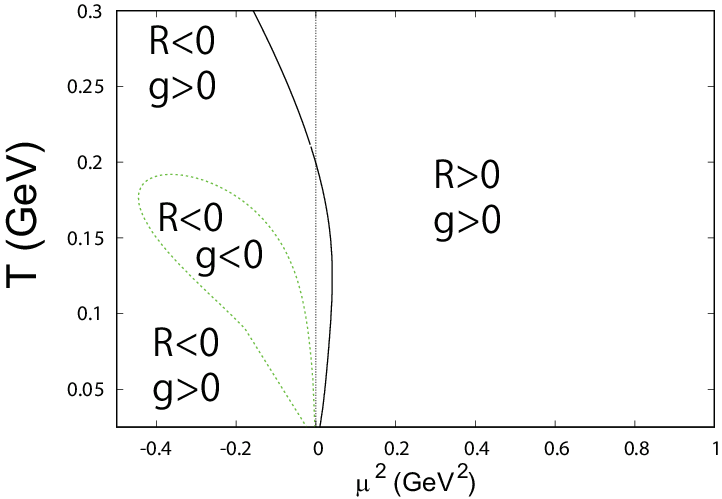}}
\caption{The curve determined by the condition $R=0$ on the $\mu^2$-$T$ plane in the HRG model without EVE.  
The meanings of the lines are the same as in Fig.~\ref{Fig_PD} and Fig.~\ref{Fig_PD_theta-T}.   
 }
 \label{Fig_PD_mu2-T}
\end{figure}

Combining Fig.~\ref{Fig_PD} (Fig.~\ref{Fig_PD_EVE}) with Fig.~\ref{Fig_PD_theta-T} (Fig.~\ref{Fig_PD_theta-T_EVE}) we obtain the whole diagram on the $\mu^2-T$ plane. 
Figure~\ref{Fig_PD_mu2-T} shows the diagram determined by the condition $R=0$ on the $\mu^2$-$T$ plane when EVE is absent. 
The curve of $R=0$ continues analytically from the real $\mu$ region to the imaginary one. 
This means that we may predict the curve of $R=0$ in the real $\mu$ region by the analytical continuation from the LQCD results in the imaginary $\mu$ region where there is no sign problem. 
In the imaginary $\mu$ region, $R$ diverges on the curve on which $g=0$.   
However, it seems that there is no counterpart of this singularity in the real $\mu$ region. 
Also note that the thermodynamic quantities themselves do not diverge on the curve.

\begin{figure}[t]
\centering
\centerline{\includegraphics[width=0.40\textwidth]{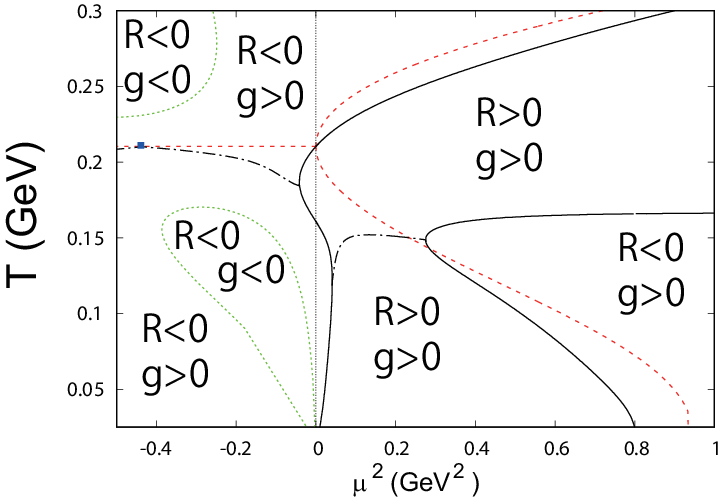}}
\caption{The solid line represents the curve determined by the condition $R=0$ on the $\mu^2$-$T$ plane in the case of the HRG model with EVE.  
The meanings of the lines are the same as in Fig.~\ref{Fig_PD_EVE} and Fig.~\ref{Fig_PD_theta-T_EVE}. 
 }
 \label{Fig_PD_mu2-T_EVE}
\end{figure}

Figure~\ref{Fig_PD_mu2-T_EVE} shows the phase diagram determined by the condition $R=0$ on the $\mu^2$-$T$ plane when EVE is present. 
The curve of $R=0$ near the $T$ axis continues analytically from the real $\mu$ region to the imaginary one and is connected to the RWL region by the curve of $|RT^3|_{\rm min}$. 
The efficiency of the HRG model with/without EVE can be checked by calculating the scalar curvature $R$ by the LQCD calculation in the imaginary chemical potential region where there is no sign problem.  
In the imaginary $\mu$ region, $R$ diverges on the curves on which $g=0$.   
It seems that there is no counterpart of this singularity in the real $\mu$ region, since this divergence of $R$ due to the vanishing of $g$ is also present in the case without EVE.

Comparing Fig.~\ref{Fig_PD_mu2-T_EVE} with Fig.~\ref{Fig_PD_mu2-T}, we see that the phase structure of the HRG model hardly depends on the EVE, when $|\mu^2|$ and $T$ are small. 
However, the phase structure depends strongly on the EVE, when $|\mu^2|$ or $T$ is large.  
In particular, the phase structure in the vicinity of the RWL point is expected to be a counterpart of the complicated phase structure in the large $\mu^2$ region.

\section{limiting temperature of baryon gas model}
\label{LTbaryon}

At $\theta =\pi$, $T_{\rm RWL}$ (=0.2103~GeV) can be regarded as the limiting temperature of the HRG model with EVE, or more precisely, the limiting temperature of the baryon gas model. 
This limitation is related with the sign change of the scalar curvature $R$. 
However, in QCD (not in the HRG model), it is expected that transition from the baryon gas to the quark matter occurs at $T=T_{\rm RW}$ which is slightly smaller than $T_{\rm RWL}$. 
On the other hand, in the region of real $\mu$, the critical point (CP) predicted by the LQCD calculation lies just below the right-lower curve of the $R=0$ criterion. 
This small disagreement indicates that the HRG model with EVE is not applicable beyond the critical point. 
As is the case of the RW transition, in the region of real $\mu$ in the QCD phase diagram, it is expected that there is true limiting temperature of the baryon gas, which is slightly lower than the temperature of the $R=0$ criterion in the HRG model with EVE.    
In this section, we investigate the limitation of the baryon gas model within the framework of the HRG model with EVE. 

\begin{figure}[t]
\centering
\centerline{\includegraphics[width=0.40\textwidth]{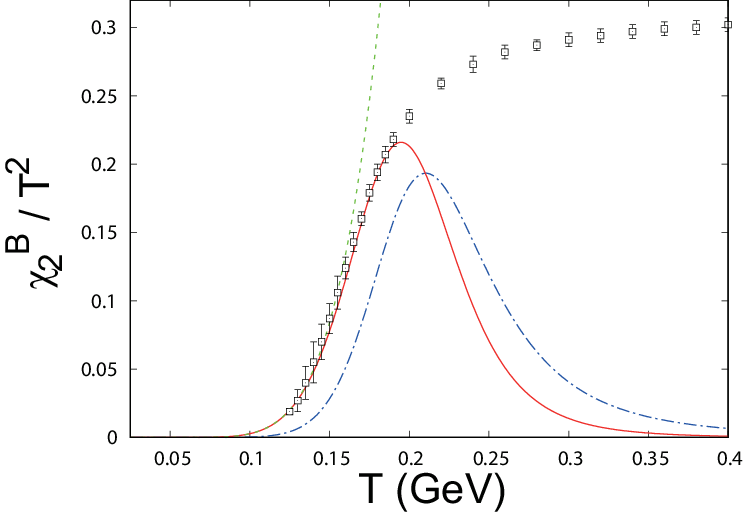}}
\caption{The solid (dotted) line shows the $T$-dependence of the dimensionless fluctuation $\chi_2^{\rm B}/T^2$ at $\mu =0$ in the HRG model with EVE  (without EVE). 
The squares with error bar show the LQCD results in Ref.~\cite{Borsanyi:2011sw}. 
The dash-dotted line shows  $T$-dependence of the dimensionless quantity  $\phi_{\gamma\gamma}/T_{\rm RWL}^3$ in the HRG model with EVE.   
}
 \label{Fig_sus}
\end{figure}

As was seen in the previous section, at $\mu =0$, $\phi_{\beta\gamma\gamma}$ vanishes at $T=T_1=T_{\rm RWL}$ where $R=0$ and $\phi_{\gamma\gamma}$ has its maximum there.  
$\phi_{22}$ is related with the baryon number fluctuation $\chi_2^{\rm B}$ according to the following relation.  
\begin{eqnarray}
{\chi_2^{\rm B}\over{T^2}}={1\over{T^2}}{\partial n_{\rm B}(T,\mu) \over{\partial \mu}}={\phi_{\gamma\gamma}\over{T^3}}. 
\label{chi2B}
\end{eqnarray}
Figure~\ref{Fig_sus} shows $T$-dependence of $\chi_2^{\rm B}/T^2$ at $\mu =0$ in the HRG models with EVE and without EVE. 
In the HRG model without EVE, $\chi_2^{\rm B}/T^2$ increases monotonically as $T$ increases. 
In the HRG model with EVE, it has its maximum at $T_{\rm max}=0.195$~GeV.  
Due to the additional factor $T^{-3}$, $T_{\rm max}=0.195$~GeV is somewhat lower than $T_{\rm RWL}$.  
We see that the results in the HRG model with EVE are consistent with those in LQCD up to $T_{\rm max}$. 
This indicates that $T_{\rm max}$ is the limiting temperature of the baryon gas model.  

\begin{figure}[t]
\centering
\centerline{\includegraphics[width=0.40\textwidth]{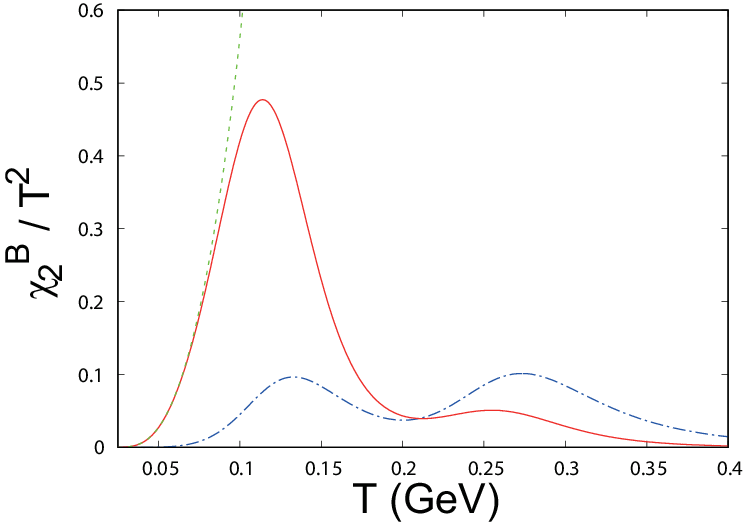}}
\caption{The $T$-dependence of the dimensionless fluctuation $\chi_2^{\rm B}/T^2$ at $\mu =0.6$~GeV in the HRG models with EVE  and without EVE.   
The meaning of the lines is the same as in Fig.~\ref{Fig_sus}.  
 }
 \label{Fig_sus_mu=600MeV}
\end{figure}

Figure~\ref{Fig_sus_mu=600MeV} shows $T$-dependence of $\chi_2^{\rm B}/T^2$ at $\mu =0.6$~GeV in the HRG models with EVE and without EVE. 
Again, in the HRG model without EVE, $\chi_2^{\rm B}/T^2$ increases monotonically as $T$ increases. 
In the HRG model with EVE,  it has two local maxima and one local minimum.  
The temperature of lower (upper) local maxima of $\chi_2^{\rm B}/T^2$ is somewhat lower than the temperature of lower (upper) local maximum of $\phi_{\gamma\gamma}/T_{\rm RWL}^3$. 
It is also seen that the lower local maximum of $\chi_2^{\rm B}/T^2$ is the global maximum. 
Hereafter, we regard the temperature of the global maximum of $\chi_2^{\rm B}/T^2$ as the limiting temperature of the baryon gas model. 

In two dimensional conformal field theory, using the renormalization group method, it was shown that the effective degrees of freedom decrease as the energy scale of the system decreases~\cite{Zamolodchikov:1986gt}. 
It is called the c-theorem. 
Inversely, it may be natural that the effective degrees of freedom increase as the energy scale such as temperature increases. 
Without EVE, the effective degrees of freedom (EDOF) of the HRG model increases as $T$ increases. 
However, EVE reduces EDOF and the dimensionless quantity $\chi_2^{\rm B}/T^2$ has its maximum at $T_{\max}$ where the two effects are balanced. 
It is unnatural that EVE overcomes the effect of increase of temperature and makes $\chi_{2}^{\rm B}/T^2$ decrease as temperature increases.
Hence, we regard the lower $T_{\rm max}$ as the limiting temperature of the HRG model with EVE.

\begin{figure}[t]
\centering
\centerline{\includegraphics[width=0.40\textwidth]{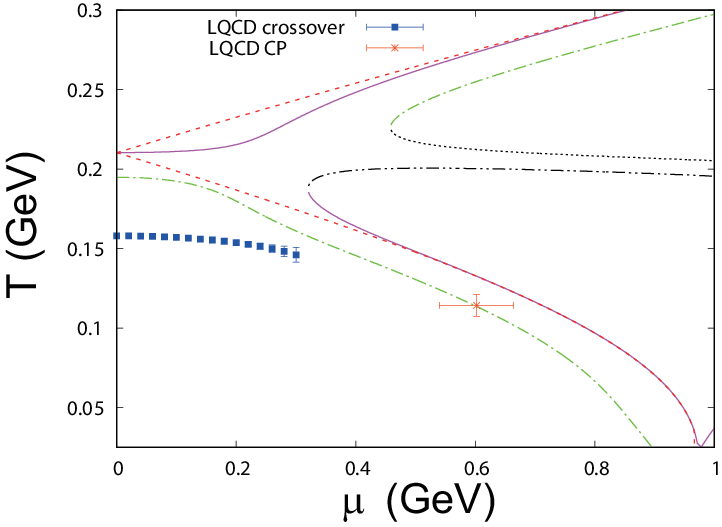}}
\caption{The dashed-dotted (solid) line shows the position of the local maximum of $\chi_2^{\rm B}/T^2$ ($\phi_{\gamma\gamma}/T_{\rm RWL}^3$) at fixed value of $\mu$ in the HRG model with EVE, while 
the dotted (dash-dot-dotted) line shows the position of its local minimum. 
The lower (upper) dashed line represents the curve determined by the condition $n_{b}=1/(2v_{\rm B}$) ($n_{a}=1/(2v_{\rm B})$). 
The meaning of the symbols is the same as in Fig.~\ref{Fig_PD}. 
 }
 \label{Fig_max_min}
\end{figure}

Figure~\ref{Fig_max_min} shows the $\mu$-dependence of $T_{\rm max}(\mu )$ ($T_{\rm min}(\mu )$) where $\chi_2^{\rm B}/T^2$ has its local maximum (minimum) for the fixed value of $\mu$ in the HRG model with EVE. 
We see that the critical point predicted by LQCD is located almost on the lower curve of $T_{\rm max}(\mu )$. 
This fact supports our assumption that the lower $T_{\rm max}(\mu )$ curve represents the limiting temperature of the baryon gas model. 
Above this temperature, quarks deconfine at least partially. 

In Fig.~\ref{Fig_max_min}
the $\mu$-dependence of $T_{\rm max}^\prime (\mu )$ ($T_{\rm min}^\prime (\mu )$) where $\phi_{\gamma\gamma}$ has its local maximum (minimum) for the fixed value of $\mu$ and   
the curve determined by the condition $n_{\rm b}=n_{\rm b0}/2=1/(2v_{\rm B})$ ($n_{\rm a}=n_{\rm a0}/2=1/(2v_{\rm B})$) are also shown.  
Note that $T_{\rm max}^\prime (\mu =0)=T_{\rm RWL}$. 
The lower (upper) curve of $T_{\rm max}^\prime$ almost coincides with the curve of $n_{\rm b}=1/(2v_{\rm B})$ ($n_{\rm a}=1/(2v_{\rm B})$) when $\mu$ is large. 
We see that the lower $T_{\rm max}(\mu )$ curve is just below the curve of $n_{\rm b}=1/(2v_{\rm B})$.  
This fact means that the limiting density $n_{\rm b}(\mu, T_{\rm max}(\mu))$ is smaller than $1/(2v_{\rm B})$. 
When $\mu$ is large, $n_{\rm B}\sim n_{\rm b}$.  
Hence, we obtain a simple empiric sufficient condition $n_{\rm B}>1/(2v_{\rm B})$ for quark deconfinement at large $\mu$, if we regard the lower $T_{\rm max}(\mu )$ as the limiting temperature of baryons.

\begin{figure}[t]
\centering
\centerline{\includegraphics[width=0.40\textwidth]{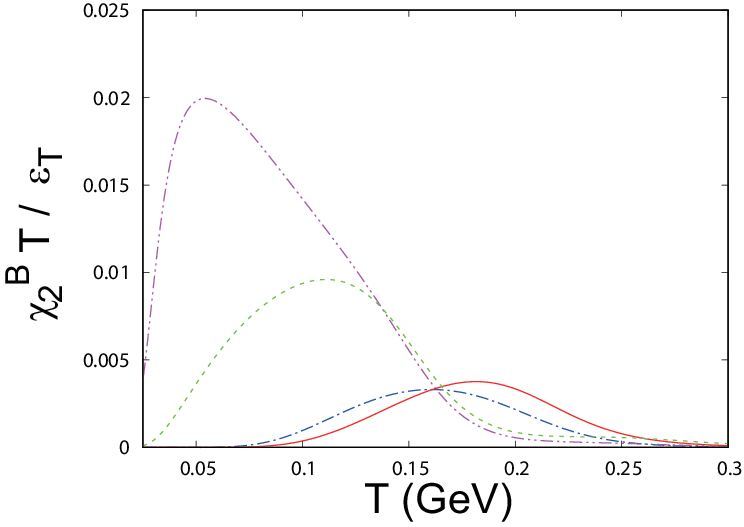}}
\caption{The solid (dashed) line shows the $T$-dependence of the dimensionless ratio $\chi_2^{\rm B}T/\varepsilon_T$ at $\mu =0$ ($\mu =0.6$ GeV) in the HRG modes with EVE. 
The dash-dotted (dash-dot-dotted) line shows the $T$-dependence of the dimensionless ratio $\phi_{\gamma\gamma}T_2^2/\phi_{\beta\beta}$ at $\mu =0$ ($\mu =0.6$~GeV). 
 }
 \label{Fig_sus2_mu=0_600MeV}
\end{figure}

\begin{figure}[t]
\centering
\centerline{\includegraphics[width=0.40\textwidth]{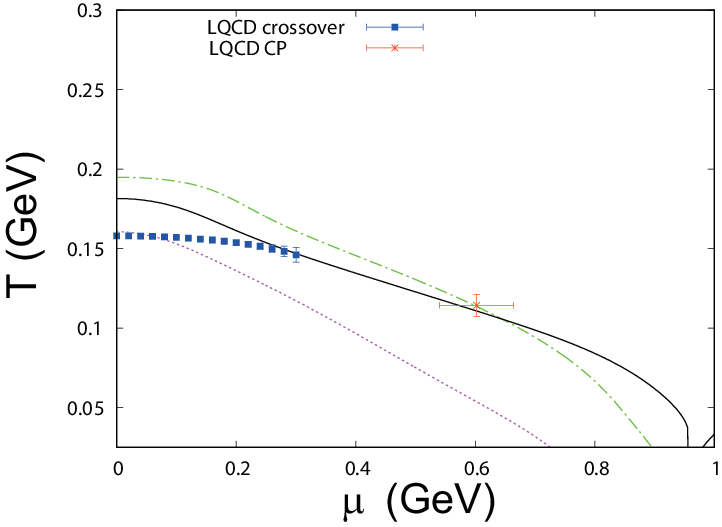}}
\caption{The solid (dotted) line shows the position of the maximum of $\chi_2^{\rm B}T/\varepsilon_T$ ($\phi_{\gamma\gamma}T/\phi_{\beta\beta}$) at fixed value of $\mu$ in the HRG model with EVE. 
The dash-dotted line shows the lower $\chi_2^{\rm B}/T^2$ maximum curve presented also in Fig.~\ref{Fig_max_min}.  
The meaning of the symbols is the same as in Fig.~\ref{Fig_PD}. 
}
 \label{Fig_max2}
\end{figure}

\begin{figure}[t]
\centering
\centerline{\includegraphics[width=0.40\textwidth]{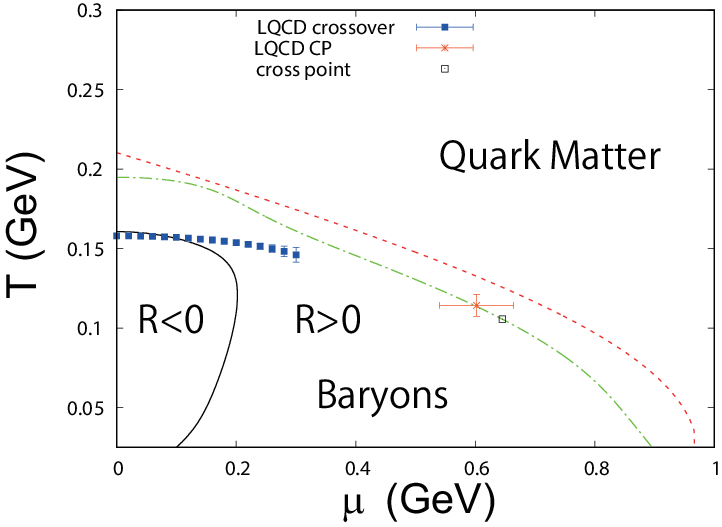}}
\caption{The phase diagram in $\mu$-$T$ plane. The square point is the cross point given by Eq.~(\ref{Eq_cross point}). 
The meaning of the lines and the other symbols is the same as in Figs.~\ref{Fig_PD} and \ref{Fig_max_min}. }
 \label{Fig_PD_final}
\end{figure}

In the HRG model with EVE, at $\mu =0$, ${\partial\over{\partial \beta}}(\phi_{\gamma\gamma}/\phi_{\beta\beta})$ vanishes at $T=T_2=0.161$~GeV and $\phi_{\gamma\gamma}/\phi_{\beta\beta}$ has its maximum there.  
$\phi_{\beta\beta}$ is related to the quantity 
\begin{eqnarray}
\varepsilon_{T}\equiv {\partial \epsilon(T,\mu)\over{\partial T}} =\beta^2\phi_{\beta\beta}+\beta\gamma\phi_{\beta\gamma}. 
\label{EpsilonT}
\end{eqnarray}
Figure~\ref{Fig_sus2_mu=0_600MeV} shows $T$-dependence of the dimensionless ratio $\chi_2^{\rm B}T/\varepsilon_{T}$ at $\mu =0$ (solid line) and $\mu =0.6$~GeV (dashed line) in the HRG model with EVE. 
At $\mu =0$ ($\mu =0.6$~GeV), the ratio has its maximum at $T_{\rm max,2}=$0.182 (0.111)~GeV which is somewhat greater than the temperature where $\phi_{\gamma\gamma}/\phi_{\beta\beta}$ has its maximum.

Figure~\ref{Fig_max2} shows the curve on which the $\chi_2^{\rm B}T/\varepsilon_T$ has its maximum in the $\mu$-$T$ plane. 
The curve meets the LQCD crossover transition line at $\mu =0.3$~GeV and crosses the curve of the $\chi_2^{\rm B}/T^2$ maximum at $\mu =0.645$~GeV.  
The cross point is given by 
\begin{eqnarray}
(\mu_{\rm CRP},T_{\rm CRP})=(0.645~{\rm GeV}, 0.106~{\rm GeV}). 
 \label{Eq_cross point}
\end{eqnarray}
It is very interesting that this cross point is very close to the critical point predicted by LQCD calculation. 
The crossover transition line is expected to end on the curve of the limitation of the baryon gas model. 
The endpoint is nothing but the critical point. 
Similarly, the curve of the $\chi_2^{\rm B}T/\varepsilon_T$ maximum in the HRG model should end at the cross point where the curve crosses the curve of the baryon gas limitation. 
Remember that, at $\mu =0.3$~GeV, the $\chi_2^{\rm B}T/\varepsilon_T$ maximum curve merges with the LQCD crossover transition line. 
Hence, this cross point may be regarded as an approximate critical point predicted by the HRG model with EVE, although the model has no explicit dynamics of the phase transition. 
Fig.~\ref{Fig_PD_final} shows the phase diagram which is predicted by our HRG model with EVE, if we assume that the curve of the $\chi_2^{\rm B}/T^2$ global maximum gives the limiting temperature of baryons. 
Above the dashed-dotted line, quarks deconfine at least partially. 
Note also that, below the dashed line, $n_{\rm b}<1/(2v_{\rm B})$ in the HRG model with EVE.

\section{Baryon radius dependence}
\label{Baryonradius}

In our model, we use the same baryon radius for all baryons. 
Hence, this baryon radius $r_{\rm B}$ is regarded as the averaged radius and its magnitude is uncertain. 
Here, we examine the $r_{\rm B}$-dependence of the results. 

\begin{figure}[t]
\centering
\centerline{\includegraphics[width=0.40\textwidth]{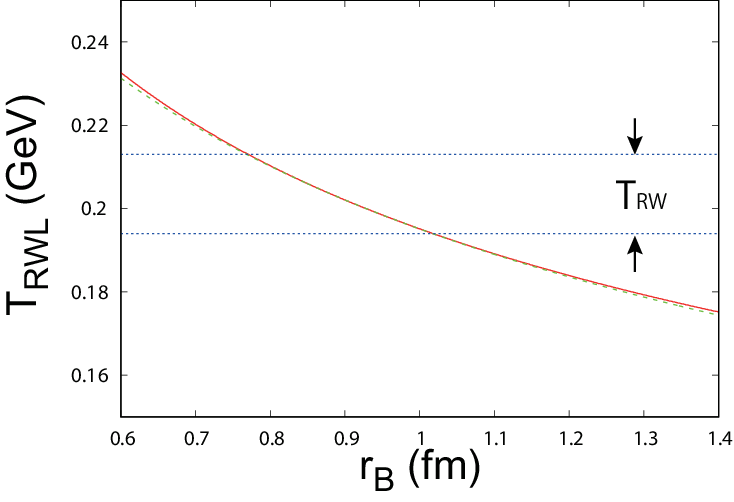}}
\caption{The solid line shows the $r_B$-dependence of the Roberge-Weiss like temperature $T_{\rm RWL}$. The dashed line shows the result obtained by using the formula (\ref{Eq_rB_TRWL}). 
The upper and lower dotted line show the maximum and minimum of the Roberge-Weiss temperature $T_{\rm RW}$ in the LQCD simulations~\cite{Bonati:2016pwz,Cuteri:2022vwk,Bonati:2018fvg}, respectively.}
\label{Fig_TRWL_RB}
\end{figure}

Fig.~\ref{Fig_TRWL_RB} shows the $r_B$-dependence of the Roberge-Weiss like temperature $T_{\rm RWL}$. 
It is seen that $T_{\rm RWL}$ decreases monotonically as $r_{\rm B}$ increases.  
The $r_{\rm B}$-dependence of $T_{\rm RWL}$ is well reproduced by the following formula.
\begin{eqnarray}
T_{\rm RWL}=0.1951{\rm GeV}/(r_{\rm B}/{\rm fm})^{1/3} 
\label{Eq_rB_TRWL}
\end{eqnarray}
Remember that $T_{\rm RWL}$ is determined by the condition $B(T_{\rm RWL})=1/v_{\rm B}$. 
It is easily seen that Eq.~(\ref{Eq_rB_TRWL}) is satisfied if $B(T)^{1/9}$ is proportional to $T$. 
In fact, in the region near $T\sim 0.2$ GeV, $B(T)^{1/9}/T$ is approximately constant. 
Hence, in this region, $B(T)$ follows the $T^9$-scaling law, the power of which is much greater than the one of the Stefan-Boltzmann limit, namely, 3. This strong power is induced by the succeeding generations of the hadron resonances. 
The higher the temperature becomes, the more significant the contributions from heavier hadrons become. 

The Roberge-Weiss like temperature $T_{\rm RWL}$ is comparable to the so-called Lambda QCD $\Lambda_{\rm QCD}$. 
In the perturbative QCD (pQCD), $\Lambda_{\rm QCD}$ represents the lower energy limit for pQCD and the running coupling diverges at $\Lambda_{\rm QCD}$. On the contrary, $T_{\rm RWL}$represents the upper energy limit for the application of the HRG model with EVE and the thermodynamic quantities diverge at $T_{\rm RWL}$ when $\theta =(2k+1)\pi$ with any integer $k$.

Since the Roberge-Weiss like singularity in the HRG model with EVE is not observed in the LQCD simulation,  
transition from baryon matter to quark matter should take place before the temperature reaches $T_{\rm RWL}$.   
Hence, $T_{\rm RWL}$ should be larger than the Roberge Weiss temperature $T_{\rm RW}$ which lies in the region $T=0.194-0.213$ GeV
~\cite{Bonati:2016pwz,Cuteri:2022vwk,Bonati:2018fvg}. 
Hence, here, we regard $T=0.194$ GeV as the minimum value of the estimated $T_{\rm RW}$. 
Figure~\ref{Fig_TRWL_RB} shows  
that $r_{\rm B}\le 1.02$ fm should be satisfied for the condition $T_{\rm RWL}\ge 0.194$ GeV. 

On the other hand, since the RWL singularity is related to the baryon instability and the RW transition, it is not natural that $T_{\rm RWL}$ is much larger than $T_{\rm RW}$. 
Hence, here we consider the cases with $r_{\rm B}=0.7$ fm and 1.0 fm. 

\begin{figure}[t]
\centering
\centerline{\includegraphics[width=0.40\textwidth]{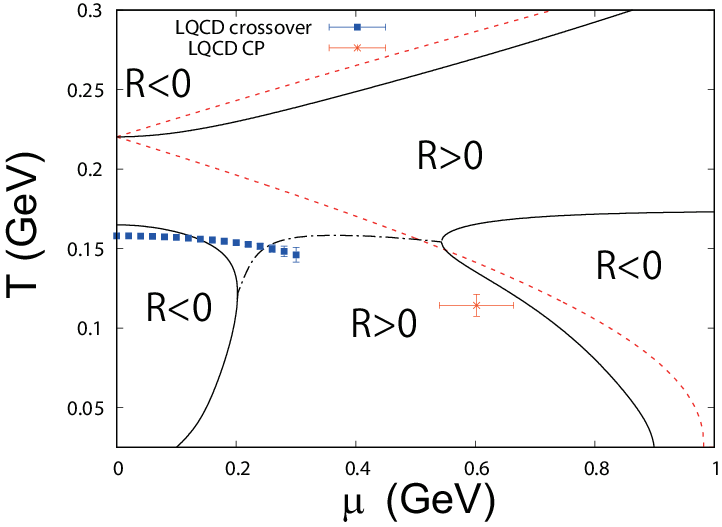}}
\caption{The solid lines represent the curves determined by the condition $R=0$ on the $\mu$-$T$ plane in the HRG model with EVE when $r_{\rm B}=0.7$ fm. 
The dash-dotted line represents the curve on which $|RT^3|$ has its local minimum for the fixed value of $\mu$ in the region where 0.202~GeV$<\mu <0.544$~GeV and $T<T_{\rm RWL}$. 
The lower (upper) dashed line represents the curve determined by the condition $n_{\rm b}=1/(2v_{\rm B})$ ($n_{\rm a}=1/(2v_{\rm B}$)). 
The meaning of the symbols is the same as in Fig.~\ref{Fig_PD}. 
 }
 \label{Fig_PD_EVE_rB=07}
\end{figure}

\begin{figure}[t]
\centering
\centerline{\includegraphics[width=0.40\textwidth]{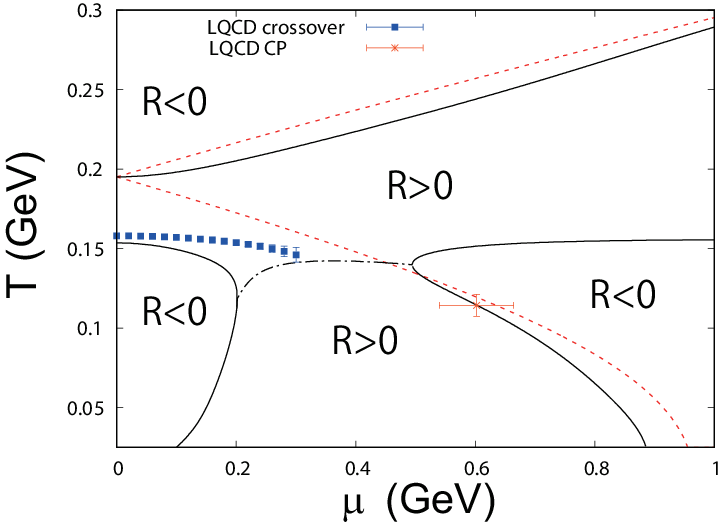}}
\caption{The solid lines represent the curves determined by the condition $R=0$ on the $\mu$-$T$ plane in the HRG model with EVE when $r_{\rm B}=1.0$ fm. 
The dash-dotted line represents the curve on which $|RT^3|$ has its local minimum for the fixed value of $\mu$ in the region where 0.200~GeV$<\mu <0.495$~GeV and $T<T_{\rm RWL}$. 
The lower (upper) dashed line represents the curve determined by the condition $n_{\rm b}=1/(2v_{\rm B})$ ($n_{\rm a}=1/(2v_{\rm B}$)). 
The meaning of the symbols is the same as in Fig.~\ref{Fig_PD}. 
 }
 \label{Fig_PD_EVE_rB=10}
\end{figure}

Figure~\ref{Fig_PD_EVE_rB=07} shows the $R=0$ condition in the HRG model with EVE when $r_{\rm B}=0.7$ fm. 
It is seen that the  $R=0$ curve near the critical point is somewhat higher than the one in the model with $r_{\rm B}=0.8$ fm. 
Similarly, the lower $R=0$ curve in the small $\mu$ region is slightly higher than the LQCD crossover line. 
However, qualitative features of the diagram hardly changed from Fig.~\ref{Fig_PD_EVE}.   

Figure~\ref{Fig_PD_EVE_rB=10} shows the $R=0$ condition when $r_{\rm B}=1.0$ fm. 
The $R=0$ curve near CP is somewhat lower than the one in the model with $r_{\rm B}=0.8$ fm and the CP locates almost on the curve.   
On the other hand, the lower $R=0$ curve in the small $\mu =0$ region is slightly lower than the crossover line. 
However, again, qualitative features of the diagram hardly changed from Fig.~\ref{Fig_PD_EVE}.  

It should be noted that the $r_{\rm B}$-dependence of $T_{\rm RWL}$ in (\ref{Eq_rB_TRWL}) is much weaker than $T_{\rm RWL}\sim 1/r_{\rm B}$ which is expected by the naive dimensional analyses. 
Thanks to the $T^9$-scaling of $B(T)$, $T_{\rm RWL}$ depends on $r_{\rm B}$ only weakly and the qualitative features of the results are robust under the moderate variations of $r_{\rm B}$.

\begin{table}[h]
\begin{center}
\begin{tabular}{|c|c|c|c|c|} \hline
 $r_{\rm B}$       &~0.7 fm~&~0.8 fm~&~1.0 fm~& LQCD \\ \hline
 $T_{\rm RWL}$   &  0.220  & 0.210 &  0.195 & 0.194$\text{--}$0.213 \\ \hline
 $T_2$ &  0.165  & 0.161 &  0.154 &  0.15808$\pm$0.00047\\ \hline
 $T_{\rm max}(0)$ & 0.203 & 0.195 &  0.182 &  0.15808$\pm$0.00047 \\ \hline
 $T_3$ &  0.135 & 0.127  &  0.115 & 0.1143$\pm$0.0069\\ \hline
 $T_{\rm max}(\bar{\mu}_{\rm CP})$ & 0.121 & 0.114 &  0.103  & 0.1143$\pm$0.0069 \\ \hline
 \end{tabular}
\caption{Summary of the $r_{\rm B}$ dependence of the critical/limiting temperature shown in GeV. 
The $T_3$ is the temperature of the lower $R=0$ curve at $\mu =\bar{\mu}_{\rm CP}=0.6021$ GeV, the average value of the chemical potential of CP in the LQCD~\cite{Shah:2024img}.   
The LQCD reference value in the second row is the value of $T_{\rm RW}$ in Refs.~\cite{Bonati:2016pwz,Cuteri:2022vwk,Bonati:2018fvg}. 
The LQCD reference values in the third and fourth rows are the temperature of the crossover line at $\mu =0$ in Ref.~\cite{Borsanyi:2020fev}, while the ones in the fifth and the sixth rows are the temperature $T_{\rm CP}$ of the critical point in Ref.~\cite{Shah:2024img}. 
}
\label{table1}
\end{center}
\end{table}

However, quantitative features of the results are somewhat changed under the variation of $r_{\rm B}$. 
Table~\ref{table1} summarizes the critical or limiting temperature in the HRG model with EVE when $r_{\rm B}=0.7,0.8,1.0$ fm. 
In all cases, the condition $T_{\rm RWL}\ge 0.194$ GeV is satisfied.  Note that $T_{\rm RWL}$ is equal to the temperature $T_1$ which is the temperature of the upper $R=0$ curve at $\mu =0$. 
We see that $T_{\rm CP}$ almost coincides with $T_{\rm max}(\bar{\mu}_{\rm CP})$ in the model with $r_{\rm B}=0.8$ fm, while it almost coincides with $T_3$ in the model with $r_{\rm B}=1.0$ fm. 
To eliminate the ambiguity of $r_{\rm B}$, more accurate determination of the Roberge-Weiss temperature $T_{\rm RW}$ is needed. 
However, since the critical point always locates below the curve of $n_{\rm b}=1/(2v_{\rm B})$ in all cases, the condition $n_{\rm B}>1/(2v_{\rm B})$
 is an empiric sufficient condition for the deconfinement of quarks at large $\mu$. 

\begin{figure}[t]
\centering
\centerline{\includegraphics[width=0.40\textwidth]{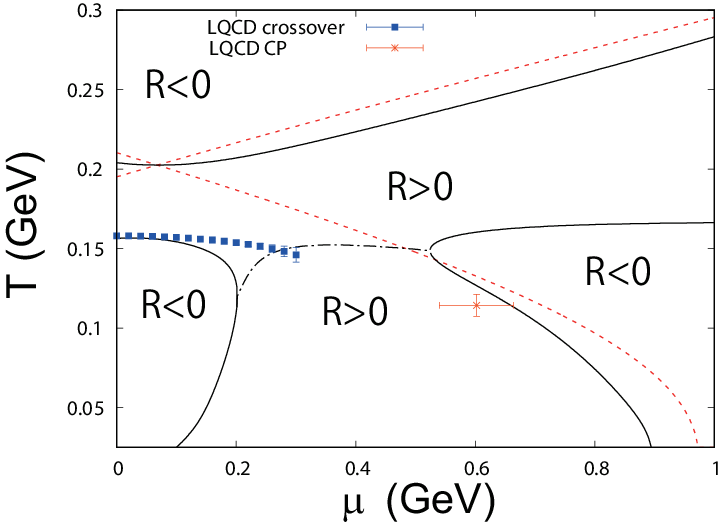}}
\caption{The solid lines represent the curves determined by the condition $R=0$ on the $\mu$-$T$ plane in the HRG model with EVE, when the baryon radius $r_{\rm b}=0.8$ fm and the antibaryon radius $r_{\rm a}=1.0$ fm.  
The dash-dotted line represents the curve on which $|RT^3|$ has its local minimum for the fixed value of $\mu$ in the region where 0.201~GeV$<\mu <0.525$~GeV and $T<T_{\rm RWL}$. 
The lower (upper) dashed line represents the curve determined by the condition $n_{\rm b}=1/(2v_{\rm b})$ ($n_{\rm a}=1/(2v_{\rm a})$) where $v_{\rm b}={4\pi\over{3}}r_{\rm b}^3$ and $v_{\rm a}={4\pi\over{3}}r_{\rm a}^3$. 
The meaning of the symbols is the same as in Fig.~\ref{Fig_PD}. 
 }
 \label{Fig_PD_EVE_rb=08_ra=09}
\end{figure}

It is expected that baryons are compressible to some extent before they collapse. 
Hence, at large $\mu$, the radius $r_{\rm b}$ of a baryon may be smaller than the radius $r_{\rm a}$ of an antibaryon.   
Figure~\ref{Fig_PD_EVE_rb=08_ra=09} shows the $R=0$ condition in the HRG model with EVE when $r_{\rm b}=0.8$ fm and $r_{\rm a}=1.0$ fm. 
It is seen that the structure of the $R=0$ curve near the critical point is not hardly changed from Fig.~\ref{Fig_PD_EVE}, since the antibaryon contribution is negligible in that region. Of course, this setting is not appropriate at $\mu =0$, since it breaks the charge symmetry at that point. 
To preserve the charge symmetry at $\mu =0$, $r_{\rm B}$ should depend on $\mu$. 
However, in such a modification, the simple pressure forms (\ref{Pb_EVE_Boltz}) and (\ref{Pa_EVE_Boltz}) are not retained. 
Similarly, introduction of species-dependent $r_{\rm B}$ in the model is also very difficult in the present formulation. 
Hence, we postpone such modifications in the future research. 

\bigskip

\section{Summary}
\label{summary}

In summary, we have studied the thermodynamic geometry in the HRG model at real and imaginary baryon chemical potential. 
In particular, we analyzed the thermodynamic scalar curvature $R$. 
In $\mu$-$T$ ($\theta$-$T$, $\mu^2$-$T$) plane, we have plotted the curves on which $R=0$. 
The main results obtained in this paper are summarized as follows. 

\bigskip

\noindent
(1) In the case of the HRG model without EVE: 
The scalar curvature $R$ at real $\mu$ is negative only when $\mu$ and $T$ are small. 
This is consistent with the previous results~\cite{Castorina:2018ayy,Castorina:2018gsx}.   
The curve of $R=0$ in the $\mu^2$-$T$ plane continues analytically from the imaginary chemical potential region to the real one. 

\bigskip

\noindent
(2) In the case of the HRG model with EVE: 
The scalar curvature $R$ at real $\mu$ is negative when $\mu$ and $T$ are small. 
This is consistent with the previous result~\cite{Castorina:2018gsx}. 
There are other regions of the negative $R$ when $T$ or $\mu$ is large.  
At $\mu =0$, $R$ vanishes when $T=T_1=T_{\rm RWL}=0.2103$~GeV and $T=T_2=0.161$~GeV.  
As is in the previous result~\cite{Castorina:2018gsx}, $T_2$ is consistent with the LQCD predicted pseudocritical temperature. 
In the wide region of imaginary $\mu$, 
$R$ is negative. 
However, at $\theta =\pi$, $R$ vanishes in the limit $T\to T_{\rm RWL}$ and there is another curve of $R=0$ in the vicinity of the RWL point. 
The sign change of $R$ is related to the singularity at the RWL point.  
One of the curves of $R=0$ in the $\mu^2$-$T$ plane continues analytically from the imaginary chemical potential region to the real one. 
As is in the previous study~\cite{Castorina:2018gsx}, the part of this curve is consistent with the LQCD crossover line when $\mu <0.1$~GeV, but deviates from it when $\mu$ increases. 
On the other hand, the LQCD predicted critical point lies just below the lower curve of $R=0$ at $\mu \sim 0.6$~GeV. 
 
\bigskip

\noindent
(3) In the wide range of the imaginary chemical potential region, the determinant $g^\prime =-g$ of the metric of the $(\theta, T)$ coordinate is negative, hence, the system is expected to be thermodynamically unstable there. Nevertheless, the analytical continuation of the scalar curvature $R$ is possible from the imaginary $\mu$ region to the real one. 

\bigskip

\noindent
(4) In the real chemical potential region, the scalar curvature $R$ does not diverge. In the imaginary chemical potential region, there are regions where the determinant $g$ of the metric is negative and $R$ diverges on the curve of $g=0$. 
However, it seems that there is no counterpart of this singularity in the real $\mu$ region. 
Note also that the thermodynamic quantities are not singular on the curve of $g=0$ in the imaginary $\mu$ region. 
 
\bigskip

\noindent
(5) The phase structure of the HRG model hardly depends on the EVE, when $|\mu^2|$ and $T$ are small. 
However, the phase structure strongly depends on the EVE, when $|\mu^2|$ and/or $T$ is large.  
In the HRG model with EVE, the phase structure in the vicinity of the RWL point is expected to be a counterpart of the phase structure in the large $\mu^2$ region. 

\bigskip

We have also investigated the limiting temperature of the baryon gas model with EVE. 
The main results obtained here are summarized as follows. 

\bigskip

\noindent
(6) At $\mu =0$, in the HRG model with EVE,  
the dimensionless quantity  
$\chi_2^{\rm B}/T^2={\partial n_{\rm B}\over{\partial \mu}}/T^2$ has its maximum when $T=T_{\rm max}=0.198$~GeV and can reproduce the LQCD result up to $T_{\rm max}$. 
We have regarded $T_{\rm max}$ as the limiting temperature of the HRG model with EVE. 
In $\mu$-$T$ plane, we have plotted the curve on which $\chi_2^{\rm B}/T^2$ has its global maximum. 
The LQCD predicted CP locates almost on the curve. 

\bigskip

\noindent
(7) At $\mu =0$, in the HRG model with EVE,  
the dimensionless quantity $\chi_2^{\rm B}T/\varepsilon_T={\partial n_{\rm B}\over{\partial \mu}}T/{\partial \varepsilon\over{\partial T}}$ has its maximum when $T=T_{\rm max,2} =0.182$~GeV.  
In $\mu$-$T$ plane, we have plotted the curve on which $\chi_2^{\rm B}T/\varepsilon_T$ has its maximum. 
At $\mu =0.3$~GeV, the LQCD predicted crossover line meets the curve of $T_{\rm max,2} (\mu )$ which crosses the curve of $T_{\max}(\mu )$ at $(\mu, T)=(0.645~{\rm GeV}, 0.106~{\rm GeV})$. 
The location of the cross point may be regarded as an approximate location of CP and is very close to the LQCD predicted CP.   

\bigskip

\noindent
(8) The lower $T_{\rm max}(\mu )$ curve in the summary (6) is just below the curve of $n_{\rm b}=1/(2v_{\rm B})$.  
This fact means that the limiting density $n_{\rm b}(\mu, T_{\rm max}(\mu))$ is smaller than $1/(2v_{\rm B})$. 
When $\mu$ is large, $n_{\rm B}\sim n_{\rm b}$.  
Hence, we obtain a simple empiric sufficient condition $n_{\rm B}>1/(2v_{\rm B})$ for quark deconfinement at large $\mu$, if we regard the global $T_{\rm max}(\mu )$ as the limiting temperature of baryons.

\bigskip

In the HRG model with EVE, the curve of $R=0$ on the $\mu$-$T$ plane is consistent with the LQCD crossover line when $\mu <0.1$~GeV. 
However, the curve deviates from the LQCD crossover line as $\mu$ increases.  
On the other hand the LQCD crossover line meets the curve of $T_{\max,2}(\mu )$ at $\mu =0.3$~GeV. 
This indicates that the nature of the LQCD crossover line changes as $\mu$ increases. 
In fact, it was pointed out that, near the critical point, the ordering density is a linear combination of the scalar density, the net baryon number density and the energy density rather than the pure scalar density itself~\cite{Fujii:2004jt}.  
The effects of the baryon number density and the energy density may be more important than the scalar density itself in the large $\mu$ region.  
This may be the reason why the HRG model can predict the CP location approximately although the model has no mechanism of the chiral phase transition.    
Further studies to investigate the nature of the transition at CP are needed. 

Our HRG model with EVE has only one nonperturbative parameter, namely the volume $v_{\rm B}$ of a baryon. 
In this paper, we treated the meson gas as the ideal gas and mainly investigated the limitation of the baryon gas.  
Apparently, the meson gas model has also its limitation. 
The effects of the interaction among the mesons and the limitation of the meson gas should be studied in future. 
Furthermore, for the studies beyond the limitation of the HRG model, we need the hybrid model which has the quark degrees of freedom as well as the hadron ones. 
The study of the thermodynamic geometry using such a hybrid model~\cite{Kouno:2023ygw,Kouno:2024cgo} is very interesting.       

In this paper, we set $r_{\rm B}=0.8$ fm except for Sec. \ref{Baryonradius} in which the $r_{\rm B}$-dependence of the results is investigated. It is shown that the qualitative features are robust under the moderate variation of $r_{\rm B}$. 
However, the quantitative features are somewhat changed. 
The baryon radius $r_{\rm B}$ is related to the Roberge-Weiss like temperature $T_{\rm RWL}$ which is related to the Roberge-Weiss temperature $T_{\rm RW}$. 
To eliminate the ambiguity of $r_{\rm B}$, more accurate determination of $T_{\rm RW}$ is needed.


\begin{acknowledgments}
H.K. thanks Hajime Aoki for useful discussions. 
This work is supported in part by Grants-in-Aid for Scientific Research from the Japan Society for the
Promotion of Science (JSPS) KAKENHI (Grant No. JP22H05112).
\end{acknowledgments}


\bibliography{ref.bib}

\end{document}